\documentstyle[epsf,epsfig]{l-aa}

\begin{document}

\thesaurus{12         
              (12.03.1;  
               09.03.1;  
	       09.04.1;  
               09.19.1)} 

\title{A conspicuous increase of Galactic contamination over CMBR anisotropies
at large angular scales}


\author{Mart\'\i n L\'opez-Corredoira}

\offprints{martinlc@iac.es}

\institute{Instituto de Astrof\'\i sica de Canarias,
              E-38200 La Laguna, Tenerife, Spain}

\date{Received xxxx; accepted xxxx}

\maketitle 

\markboth{Galactic microwave anisotropies}{M. L\'opez-Corredoira}

\begin{abstract}

New calculations of the Galactic contamination over microwave
background radiation anisotropies are
carried out. On one hand, when a frequency-dependent 
contrast of molecular clouds with respect to the Galactic background of the 
diffuse interstellar medium is taken into account, the anisotropic amplitude produced by Galactic 
dust is increased with respect to previous calculations and this is
of the same order as that of the data from the observations. 
On the other hand, if we take into account rotational dust emission, 
for instance, a frequency independence of anisotropies in the microwave 
range may be obtained.

This leads to the possibility that under some particular,
but not impossible, conditions all the microwave background
radiation anisotropies may be due to 
Galactic foregrounds rather than cosmological in origin. Moreover,
a suspected coincidence between the typical angular sizes of the
microwave background radiation anisotropies and those of nearby
molecular clouds makes more plausible the hypothesis of a purely Galactic 
origin for these anisotropies. It is also argued that
the correlation among structures at different frequencies, the comparison
of the power spectrum at different frequencies and the galactic latitude
dependence of the anisotropies are not yet proofs in favour of either a 
cosmological or Galactic origin.

   \end{abstract}

\section{Introduction}

Various observations of the cosmic microwave
background radiation (CMBR) have led to the claim that it is anisotropic
(see reviews Readhead \& Lawrence 1992; White, Scott \& Silk 1994).
Measurement of these anisotropies
provides information on structural formation,
inflation, quantum gravity, topological defects (strings, etc.),
dark matter type and abundance, the determination of $H_0$,
$\Omega $, $\Lambda $, the geometry and dynamics of the Universe, the thermal
history of the Universe at the recombination epoch, etc. A knowledge of these 
anisotropies is very important
for discriminating among different cosmological models as well
as for measuring certain related parameters.

Unfortunately, this information is contaminated by several effects
which are unrelated to cosmology, and which are either extragalactic 
or Galactic in origin. Extragalactic sources of such contamination may be
gravitational scattering due to inhomogeneities in 
the matter distribution of superclusters 
(Fukushige, Makino \& Ebisuzaki 1994; Suginohara, Suginohara \& 
Spergel 1998), inhomogeneous reionization (Knox, Scoccimarro \& Dodelson 1998),
possible radiative decay of massive tau neutrinos (Celebonovic, Samurovic 
\& Circovic 1997), the inverse Compton effect, temporal variations of the 
gravitational potential, etc. This paper will not analyse such extragalactic
sources of anisotropy but will concentrate instead on those coming from our 
own Galaxy, particularly at large angular separations ($\theta > 2^\circ $).

A summary of possible contributions from our own Galaxy to the anisotropies 
has been studied, for instance, by Bennett et al. (1992) and has led to the 
conclusion that there are three possible emission mechanisms: synchrotron, 
free-free and dust emission. The calculation of anisotropies originating 
in our Galaxy was carried out on the basis of observations from the 
{\em COBE}-DIRBE, which were extrapolated 
to other frequencies, and it was
concluded that Galactic contamination is negligible. 
The paper of Bennett et al. was optimistic. However, not all authors think the
same. Masi et al. (1990) pointed out that dust may provide 
an important contribution to the anisotropies in the microwave region.
Banday \& Wolfendale (1991) predicted that 
dust contamination is quite important except in some
regions away from the cirrus. Hence, whether or not the dust contribution 
is negligible seemed to be a point of contention by the beginning of nineties.

Recently, clear evidence of correlations between far-infrared maps
(which trace Galactic dust) and microwaves has been presented (Kogut et al.
1996a, de Oliveira-Costa et al. 1997; de Oliveira-Costa et al. 1998). 
In the light of these works it is difficult to deny the influence of
the Galaxy on the microwave background. Even though the source
of the contamination remains unknown, the 
 correlation between the Galaxy and {\it COBE}-DMR data remains an
observational fact.
Another recent paper written by Pando, Valls-Gabaud \& Fang (1998) points 
out the non-Gaussian nature of the CMBR over large angular scales of anisotropies
against standard cosmological predicitions, and this might be due to an
excess of foreground contamination.

In this paper, an explanation is offered for these last observational facts. 
Two new elements are added to previous works for the 
calculation of Galactic contamination: a frequency-dependent contrast 
of molecular clouds with respect to the Galactic background of the diffuse 
interstellar medium; and the existence of a recently noticed source of
emission: rotational emission by dust grains (Draine \& Lazarian
1998a). These lead to a non-negligible and possibly unique
source of anisotropies due to our Galaxy.

\section{How large is the Galactic contamination?}

The existence of flux anisotropies due to Galactic clouds
is undeniable. In the far infrared, the presence
of the molecular clouds is associated with ``infrared cirrus''.
There is a good correlation between infrared cirrus and atomic
hydrogen gas (Boulanger, Baud \& van Albada 1985; Burton \& Deul 1987;
Boulanger \& P\'erault 1988; Schlegel, Finkbeiner \& Davis 1998)
and the fluctuations of the hydrogen column densities within the nearest 75 to
100 pc of the Sun are occasionally  about $5\times 10^{19}$ cm$^{-2}$
(Frisch \& York 1986), so this implies that fluctuations due to
nearby clouds must be detected to some extent in the microwave region.

Most of the clouds are confined in the Galactic plane
but there are few which can be observed at high
galactic latitudes, and which are close to us (Blitz 1991).
Those responsible for large angle correlations would need to have
a large size (i.e. giant molecular clouds). 
So, some microwave background radiation anisotropies (MBRAs) may be due to
inhomogeneities in the density distribution of the local interstellar
medium. 

Dust associated with these giant clouds exists in three different stages
associated with molecular, neutral and ionized hydrogen 
(Sodroski et al. 1997), whose total emission gives a continuum
spectrum proportional to the column density in the line of
sight. Column density fluctuations lead to intensity fluctuations.
Dust anisotropies would be due to its non-homogeneous
distribution in molecular clouds or its equivalent for other
types of Galactic emission. 

Dark matter around spiral galaxies in the form of cold
gas, essentially in molecular form and rotationally supported, was
purposed by Pfenninger, Combes \& Martinet (1994) and
Pfeninger \& Combes (1994); this
cold gas in molecular form was also held to contribute at microwave wavelengths
(Schaefer 1994; Schaefer 1996). The gas would be at temperatures close 
to $3$ K and composed of small clumpuscules of size $30$ AU.
It is claimed (Combes \& Pfenniger 1997; Schaefer 1996) 
that part of the radiation observed by {\em COBE}-FIRAS is cold gas 
instead of dust. It might be in the outer disc and difficult,  
although not impossible, to detect. Indeed,
Combes \& Pfenninger (1997) propose  techniques for detecting it. 
Though this cold gas may be a source of anisotropies, it would be
only for very small scales since the emission is high enough just for
very high densities ($n(H_2)\sim 10^{18}$ cm$^{-3}$) inside the small
clumpuscules. Thence, cold molecular gas emission will not be further
considered.

In following subsections, the expected emission from
our Galaxy will be compared with observations in the microwave region.

\subsection{MBRAs amplitude}

\begin{figure}
\begin{center}
\mbox{\epsfig{file=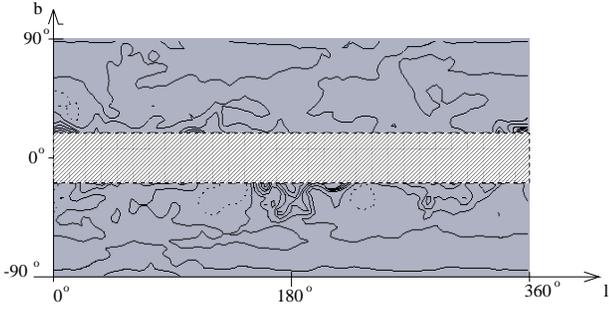,height=4cm}}
\end{center}
\caption{The {\em COBE}-DIRBE flux map at 240 $\mu $m after subtraction 
of a smooth component of the Galactic
background as explained in the text. The contours are from --20 to +20
MJy sr$^{-1}$ in steps of 4 MJy sr$^{-1}$.}
\label{Fig:DIRBE240_g_3}
\end{figure}

\begin{figure}
\begin{center}
\mbox{\epsfig{file=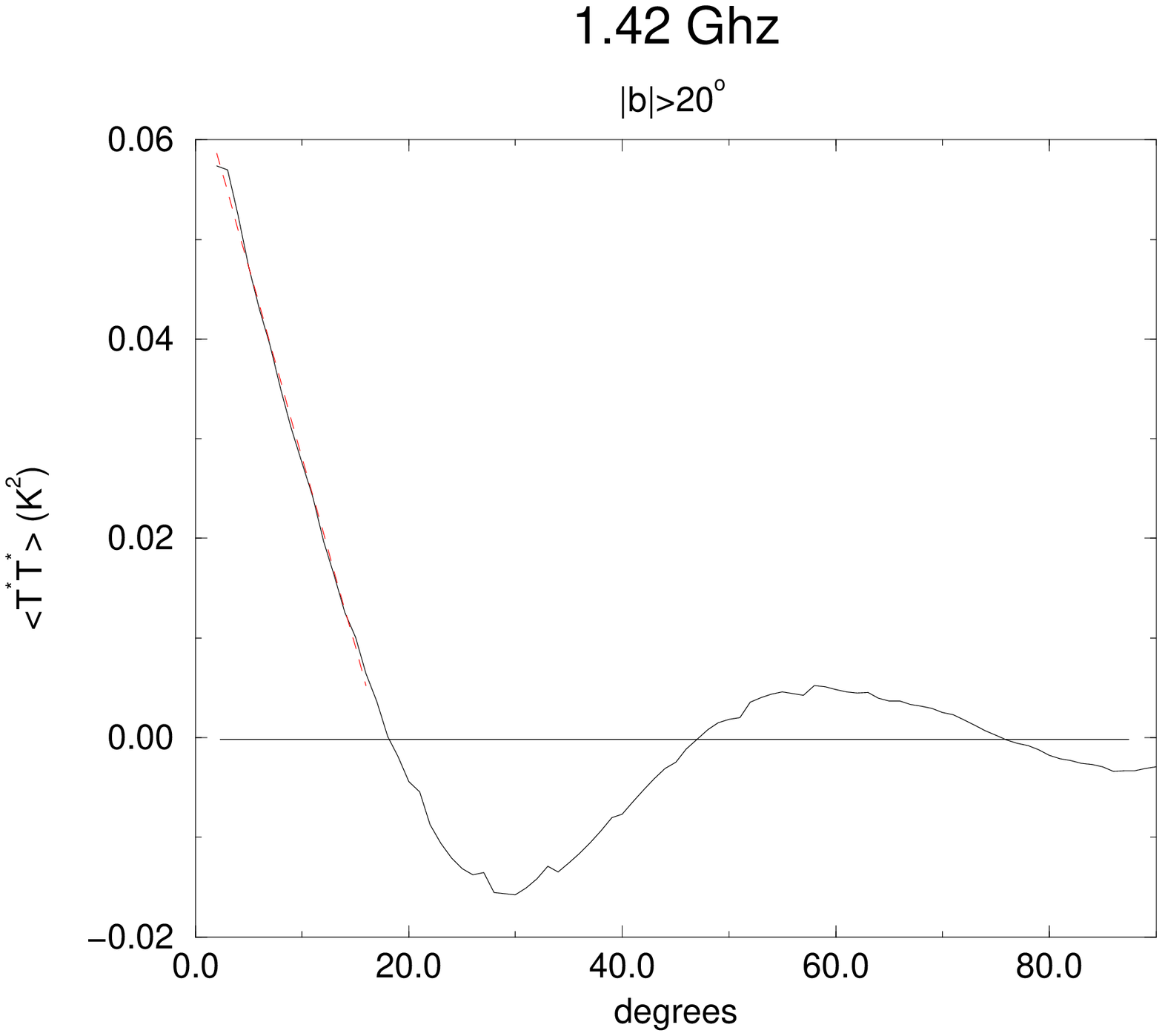,height=7cm}}

a)

\end{center}
\caption{Angular correlation function for different surveys at $|b|>20^\circ $:
a) 1.42 GHz (the Max Planck Institute for Radio Astronomy, Reich 1982;
Reich \& Reich 1986), the dashed line is a fit to
$0.0663-0.00382\theta ({\rm deg.})$ K$^2$; b) 240 $\mu $m ({\em COBE}-DIRBE, 
Boggess et al. 1992), the dashed line is a fit to $2.13\times 10^{-8}
\theta ({\rm deg.})^{-1.20}$ K$^2$.}
\label{Fig:corr}
\end{figure}

The two-point correlation function is:

\begin{equation}
\langle T^*T^* \rangle (\theta )=\int d\Omega \int _{|\Omega'-\Omega|=\theta} d\Omega ' 
T^*(\Omega )T^*(\Omega ') 
\end{equation},
where $T^*$ is the antenna temperature\footnote{Antenna 
temperatures, $T$, were obtained
from the intensity maps at 1.42 GHz and 240$\mu $m 
multiplied by the factor $\frac{c^2}{2k\nu ^2}$, 
where $c$ is the light speed and $k$ is the Boltzmann constant.}, once the 
average flux is subtracted\footnote{A background depending on Galactic
coordinates is removed to eliminate the flux variation due to its
smooth gradient. Therefore, we measure only the correlations due to 
flux anisotropies with regard this average background.
I have used bidimensional ``spline3'' functions of order 3 in both 
coordinates and
with crossed terms to fit the background in the selected regions
regions. This was done by means of the IMSURFIT task of IRAF.
These regions are selected in off-plane regions ($|b|>20^\circ$).
See the map at 240$\mu $m with the background subtracted in Figure 
\ref{Fig:DIRBE240_g_3}.
For 1.42 GHz, since the whole sky was not covered, 
only available regions with $|b|>20^\circ $ were
used ($\delta >-19^\circ $). The zodiacal component is not removed
at 240 $\mu $m as it is negligible at these frequencies (Reach et al. 1995).},
the results are those shown in Fig. \ref{Fig:corr} and \ref{Fig:corr_cross}. 

\begin{figure}
\begin{center}
\mbox{\epsfig{file=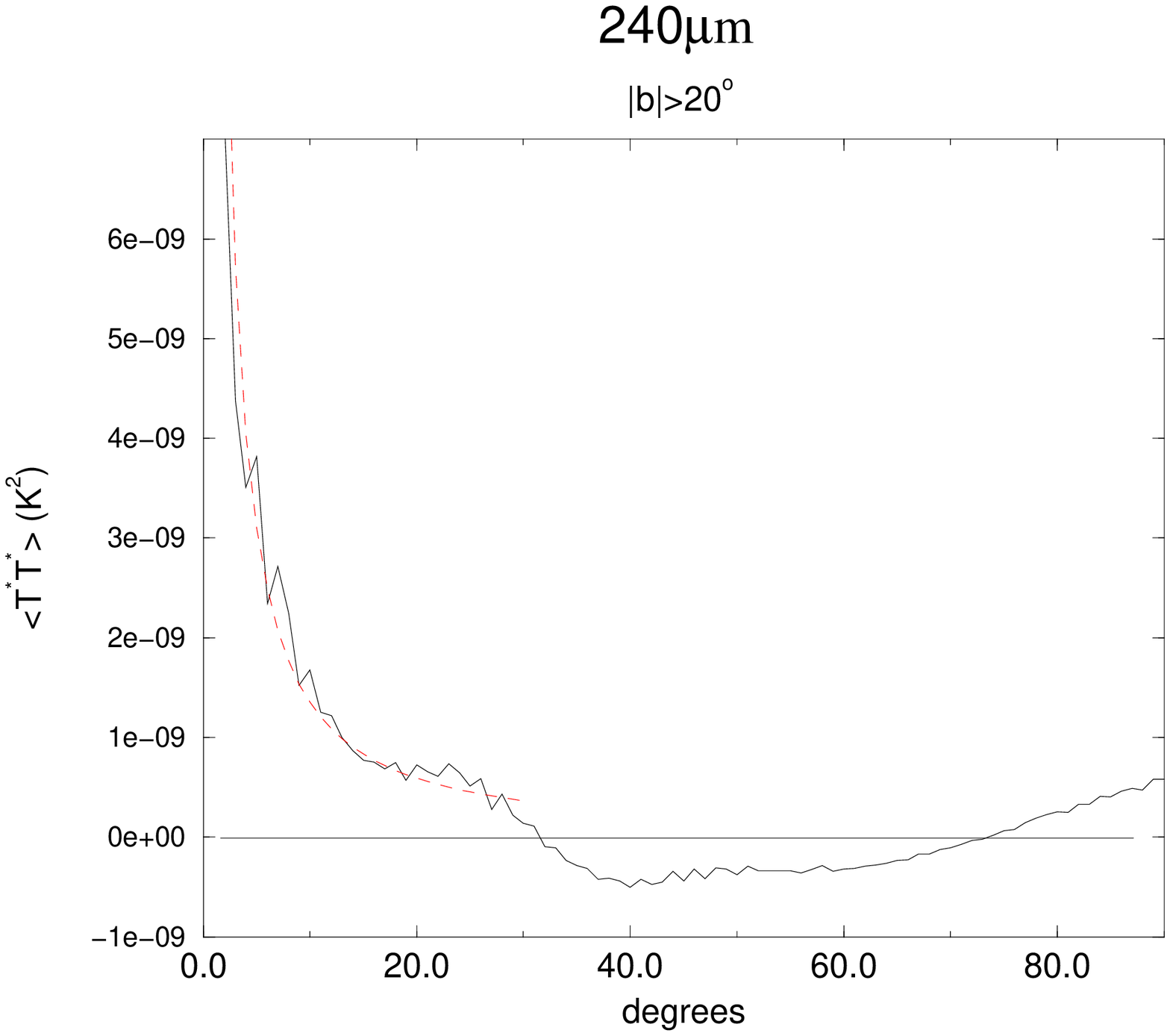,height=7cm}}

Fig. \ref{Fig:corr} b)
\end{center}
\end{figure}

\begin{figure}
\begin{center}
\mbox{\epsfig{file=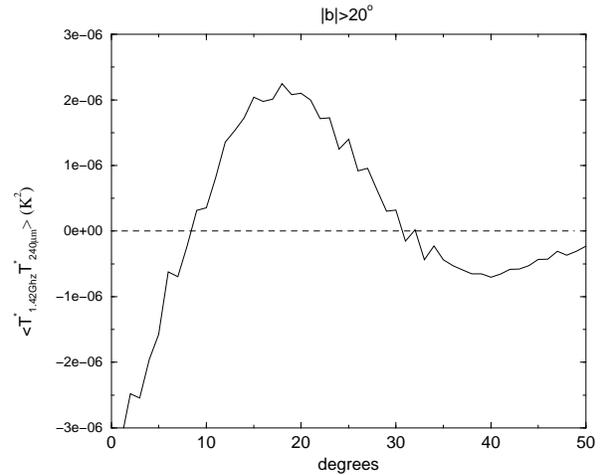,height=7cm}}
\end{center}
\caption{Cross correlation function for two surveys:
1.42 GHz (Max Planck Institute for Radio Astronomy, Reich 1982;
Reich \& Reich 1986) and 240 $\mu $m ({\em COBE}-DIRBE, Boggess et al. 1992).}
\label{Fig:corr_cross}
\end{figure}

In order to calculate the effects of the Galaxy on MBRAs, 
extrapolation of the measured two-point correlation functions 
(Fig. \ref{Fig:corr}) at 1.42 GHz due to synchrotron emission 
and at 240$\mu $m due to continuum dust emission must be carried out. 
Free-free emission is not considered here in order to simplify the calculations
and because we have no information concerning it (Smoot 1998). In any case, if
it is proven than dust or synchrotron emission or both are high enough 
then adding free-free emission would increase Galactic anisotropies, never
decrease.

\begin{figure}
\begin{center}
\mbox{\epsfig{file=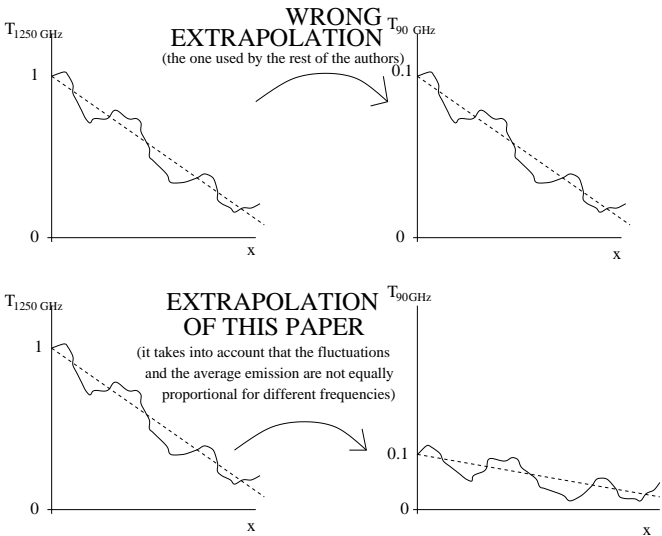,height=7cm}}
\end{center}
\caption{Example of extrapolations for Galactic dust emission. 
The main difference between the extrapolation of expression 
(\protect{\ref{ext2}}) and those derived in  work prior to
this paper is that the latter do not take into account the frequency
dependence of the proportionality
between the amplitude of the fluctuation and that of the average flux.}
\label{Fig:explica}
\end{figure}

At any frequency in the
microwave range, two-point correlation functions must be evaluated by 
extrapolating both effects independently and summing them up according to:

\[
\langle T(\nu )T(\nu ) \rangle =
\langle (T_{\rm synchr}(\nu )+T_{\rm dust}(\nu ))
\]
\[ \times(T_{\rm synchr}(\nu )
+T_{\rm dust}(\nu ))\rangle=
\langle T_{\rm synchr}(\nu )T_{synchr}(\nu ) \rangle 
\]\begin{equation}+
\langle T_{\rm dust}(\nu )T_{\rm dust}(\nu )\rangle +
2\langle T_{\rm synchr}(\nu )T_{\rm dust}(\nu )\rangle
\label{corrtotal}
.\end{equation}

Each of these terms may be calculated by means of an extrapolation
of  previously measured correlations. I will calculate the contribution
due to the local Galactic flux fluctuations with respect the average
flux, i.e. $<T^*T^*>$, and not that due to the smooth variation of the
dependence of this average flux with Galactic coordinates.
Usually, the extrapolation is carried out by multiplying each of
the pixels of the map at $\nu _0$ by the mean amplitude variation
of the flux ($\langle T(\nu)/T(\nu _0)\rangle $) and calculating
the correlations of the new map at frequency $\nu $. 
This is equivalent to multiply the
two-point correlation function by a factor 
$\langle T(\nu)/T(\nu _0)\rangle ^2$. I will go one step further
and will take into account that the local fluctuations do not vary
proportionally to the average flux (Fig. \ref{Fig:explica});
therefore, the mean amplitude
of the two-point correlation function is affected by a
second factor. As we will see later, this variation is due to different
effective dust temperatures between the Galactic clouds which produce the
fluctuations and the Galactic medium that contributes to the average flux.
The shape of two-point correlation function is also expected to vary, 
as will be explained in \S \ref{.diffreq}. However,
since we  want to estimate only the order of magnitude
of the correlation amplitude in the microwave region, 
we will not take this effect into account at this stage. 
The effects of the shape variation of the two-point correlation function
will be analysed in \S \ref{.shapevar}. Hence, extrapolations 
are calculated according to

\[
\langle T_{{\rm synchr}}(\nu )T_{{\rm synchr}}(\nu )\rangle (\theta )=
\left \langle \frac{T_{{\rm synchr}}(\nu )}
{T_{{\rm synchr}}(1.42\ {\rm GHz})}\right \rangle ^2
\]\[ \times
\left \langle\frac{\left(\frac{\Delta T_{{\rm synchr}}(\nu )}{T_{{\rm synchr}}(\nu )}\right)}
{\left(\frac{\Delta T_{{\rm synchr}}(1.42\ {\rm GHz})}
{T_{{\rm synchr}}(1.42\ {\rm GHz})}\right)}\right \rangle ^2
\]\begin{equation}\times
\langle T^*_{{\rm synchr}}(1.42\ {\rm GHz})T^*_{{\rm synchr}}
(1.42\ {\rm GHz})\rangle (\theta )
,\end{equation}

\[
\langle T_{\rm dust}(\nu )T_{\rm dust}(\nu )\rangle (\theta )=
\left \langle \frac{T_{\rm dust}(\nu )}
{T_{\rm dust}(240\ \mu{\rm m})}\right \rangle ^2
\]\[\times
\left \langle \frac{\left(\frac{\Delta T_{\rm dust}(\nu )}{T_{\rm dust}(\nu )}\right)}
{\left(\frac{\Delta T_{\rm dust}(240\ \mu{\rm m})}
{T_{\rm dust}(240\ \mu{\rm m})}\right)}\right \rangle ^2
\]\begin{equation}\times
\langle T^*_{\rm dust}(240\ \mu{\rm m})T^*_{\rm dust}(240\ \mu{\rm m})
\rangle (\theta )
\label{ext2}
,\end{equation}

\[
\langle T_{{\rm synchr}}(\nu )T_{\rm dust}(\nu )\rangle (\theta )
\]\[=
\left \langle \left ( \frac{T_{{\rm synchr}}(\nu )}
{T_{{\rm synchr}}(1.42\ {\rm GHz})}\frac{T_{\rm dust}(\nu )}
{T_{\rm dust}(240\ \mu{\rm m})}\right )\right \rangle
\]\[ \times
\left \langle \left (\frac{\left(\frac{\Delta T_{{\rm synchr}}(\nu )}
{T_{{\rm synchr}}(\nu )}\right )}
{\left(\frac{\Delta T_{{\rm synchr}}(1.42\ {\rm GHz})}
{T_{{\rm synchr}}(1.42\ {\rm GHz})}\right)}
\frac{\left(\frac{\Delta T_{\rm dust}(\nu )}{T_{\rm dust}(\nu )}\right)}
{\left(\frac{\Delta T_{\rm dust}(240\ \mu{\rm m})}
{T_{\rm dust}(240\ \mu{\rm m})}\right)}\right )\right \rangle
\]\begin{equation}\times 
\langle T^*_{{\rm synchr}}(1.42\ {\rm GHz})T^*_{\rm dust}
(240\ \mu{\rm m})\rangle(\theta )
,\end{equation}
so we have to multiply three factors for any of the correlations: the
first factors ($\langle T(\nu)/T(\nu _0)\rangle $) take into account 
the variation of the mean Galactic background, the second factors 
($\langle \frac{\Delta T}{T}(\nu )/\frac{\Delta T}{T}(\nu _0)\rangle $) for the variation 
of relative fluctuations at different frequencies and the third factors 
($<T^*T^*>(\nu _0, \theta)$) are 
respectively the correlations measured in Fig. \ref{Fig:corr} a), 
\ref{Fig:corr} b) and \ref{Fig:corr_cross}. The three factors can
be separated because of the assumption of the independence of $\theta $ for
the first and the second factor.

\begin{description}

\item[First factors:]

The antenna temperature due to synchrotron emission is

\begin{equation}
T_{{\rm synchr}}(\nu ,|b|>20^\circ )= 0.6\left(\frac{\nu }{1420\ {\rm MHz}}\right)^{\beta}
\ \ {\rm K},\end{equation}
where $\beta \approx -3$ (Davies, Watson \& Guti\'errez 1996).
See the contribution in Figure \ref{Fig:factor1}.

The dust temperature has two contributions: thermal emission,
which is detected at far infrared wavelengths (this is the emission
detected at 240 $\mu $m); and the microwave 
rotational emission due to small spinning 
grains (Draine \& Lazarian 1998a). The first contribution is widely
accepted while the second is still usually not included in the
calculation of the Galactic contribution to the anisotropies in the microwave
range.

The mean thermal dust emission is given 
(Fig. \ref{Fig:factor1}) by Reach et al. (1995):  

\[
T_{\rm dust}(\nu ,|b|>20^\circ )=\left(\frac{\nu }{900\ {\rm GHz}}\right)^2 
[1.74\times 10^{-5}
\]\[\times
B(\nu ,T=17.72\ {\rm K})+1.23\times 10^{-4}
B(\nu ,T=6.75\ {\rm K})]
\]\begin{equation}\times
\frac{c^2}{2k\nu ^2} \ \ {\rm K}
\label{dust>20}
,\end{equation}
where $B$ is the blackbody radiation intensity, $c$ the light speed, $k$
the Boltzmann constant and $\frac{c^2}{2k\nu ^2}$ is the factor
for converting intensities into antenna temperatures. This is an average
law for $|b|>20^\circ $; as shown by Reach et al. (1995), 
there may be important deviations from this law for some regions.

The second term of eq. (\ref{dust>20}) may be due to
very cold clouds; their average temperature would be 6.75 K if dust
emission were proportional to $\nu ^\alpha B(\nu, T)$ with $\alpha =2$,
although this would be somewhat different if $\alpha $ had another value.
Expression (\ref{dust>20}) is just a fit to the observational 
{\it COBE}-FIRAS data, not a model, and their parameters may
not have any direct physical meaning.
It was argued by Lagache et al. (1998) that an alternative hypothesis
to the existence of these very cold clouds is a Cosmic Far Infrared
Background due to distant galaxies of 26 $\mu $K in the range between
400 $\mu $m and 1000 $\mu $m (Puget et al. 1996)\footnote{This 
extragalactic background
is obtained after subtracting the CMBR,
zodiacal emission and Galactic dust emission from {\em COBE}-FIRAS data, 
making an extrapolation of the HI column
density data with an emission law $\nu ^2 B(\nu, T=17.5 K)$.
Obviously, if it is assumed at first that there are not very cold clouds
then the conclusion must be that there are no very cold clouds, and this is what
these authors find, attributing the remnant emission to an extragalactic
source. This remnant emission is quite isotropic and
led Puget et al. (1996) and Boulanger et al. (1996) to think that this 
component does not belong to the Galactic disc; however, 
it is difficult to tell what really happens because of low
signal-to-noise at high Galactic latitudes and neither does
Galactic cirrus have a clear position dependence at high latitudes (see \S
\ref{.posdep}); thus, I think these local sources are not excluded from 
containing very cold clouds. Moreover,
the extrapolation of HI column density to far infrared emission 
is also dangerous since the correlation is not rigorous (see
\S \ref{.diffreq}). HI is a dust tracer but not a perfect one and
it may be not appropriate at all to find out information about the origin
of some small remnants of {\it COBE}-FIRAS which do not correlate with it.
From all these considerations, I think it is risky to place
the origin of some very far infrared radiation (400 to 1000 micron) outside our
Galaxy.}.  Regardless whether or not this hypothesis is true,  for our purposes
the extragalactic emission can be added to the average flux of the Galaxy.
The important thing, as will be seen later, is that the fluctuations of the
flux are due to local clouds, and that the average temperature of the sources
which are origin of the fluctuations is less than that corresponding
to the diffuse emission of the Galaxy (plus any extragalactic background).

The rotational emission is the predicted model by
Draine \& Lazarian (1998a; preferred model: A), Fig. \ref{Fig:factor1} in this
paper, taking into account a hydrogen column density
$N(H)=3.9\times 10^{20}[cosec(b)-0.17]$,
which includes neutral and ionized hydrogen (Heiles 1976; Reynolds 1991), 
and averaged over $|b|>20^\circ $.

\begin{figure}
\begin{center}
\mbox{\epsfig{file=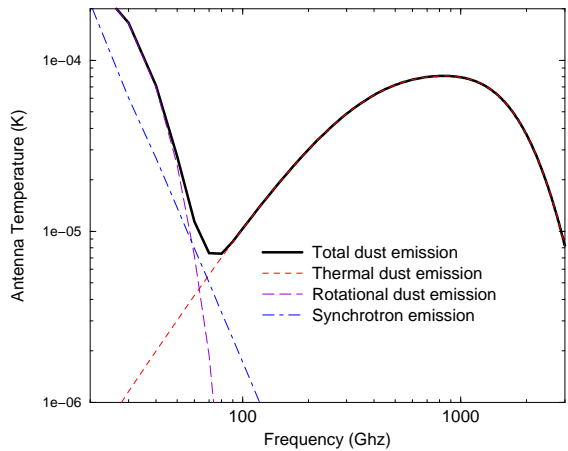,height=7cm}}
\end{center}
\caption{Mean Galactic background emissions for $|b|>20^\circ$.}
\label{Fig:factor1}
\end{figure}

\item[Second factors:]
I stress that, as far as I know, 
all authors who make the extrapolation of correlations
set the second factor equal to unity for all frequencies, i.e.
the relative fluctuations for dust or synchrotron do not depend on
the frequency (for example, Banday \& Wolfendale 1991; Bennett et al. 1992;
Guarini, Melchiorri \& Melchiorri 1995; Reach et al. 1995;
Femen\'\i a et al. 1998). This is a very poor approximation for dust,
especially when the extrapolation is to one order of magnitude
for lower frequencies, as it is in this case.
I think this poor approximation is the main reason why
dust anisotropies were considered negligible in the past.

The reason why I think this is a bad approximation for dust is that the
mean dust emission intensity follows a different continuum spectrum
from that of the local clouds (infrared cirrus) providing 
fluctuations at scales of
several degrees. When the dust emission is modelled by $I\propto
\nu ^\alpha B(\nu, T)$ (with $B$ the blackbody radiation),
$\alpha $ varies from place to place (Tegmark 1998):
it is smaller for molecular clouds (Mathis 1990; Schloerb, Snell \&
Schwartz 1987; Meinhold et al. 1993) or higher latitudes
(Banday \& Wolfendale 1991; Reach et al. 1995)  than
that for the mean intensity (Matsumoto et al. 1988; Wright et al. 1991;
Reach et al. 1995). Moreover, the mean intensity includes emission from
star-forming regions and mass-loss stars in which the dust is hotter
than the diffuse or clouds dust (Draine 1994).
The $\alpha $ spectral index depends on
the grain properties (Wright 1993) varying from 1 to 3.5 
(Banday \& Wolfendale 1991).
It is $\approx 2$ for graphite, $\approx 1.5$ for silicates
and $\approx 1$ for layer-lattice materials, such as amorphous carbon
(Draine \& Lee 1984).
The difference is also related to the different distribution
of dust-grain sizes in the diffuse interstellar medium and in molecular clouds.
The rate of small grains is larger in the diffuse interstellar
medium than in the clouds and the temperatures are different
implying that the peak emission from clouds occurs at 
different wavelength as compared with that of the diffuse medium 
(Greenberg \& Li 1996).

These second factors are unknown because the anisotropies are not sufficiently
explored for wavelengths longer than 240 $\mu $m. 
In order to give a rough estimate, a simplifying assumption can be made: 
the Galactic contribution is the sum of the local clouds---infrared cirrus---
($T_{\rm c}$)
plus a diffuse Galactic background ($T_{\rm d}$), where all the anisotropies 
at angular scales of several degrees are due to anisotropies in the local clouds 
distribution ($\Delta T_{\rm d}=0$).
Hence,

\begin{equation}
T_{\rm dust}=T_{\rm dust,d}+T_{\rm dust,c}
\label{dustdc}
\end{equation}

\begin{equation}
\frac{\Delta T_{\rm dust}(\nu )}{T_{\rm dust}(\nu )}=
\frac{\left(\frac{\Delta T_{\rm dust,c}(\nu )}{T_{\rm dust,c}(\nu )}\right)}
{1+\frac{T_{\rm dust,d}(\nu )}{T_{\rm dust,c}(\nu )}}
.\end{equation}

The numerator of this expression is constant for all frequencies
because the fluctuations of the temperature due to clouds are
proportional to their density fluctuations for any frequency.
However, the denominator is not constant. 
The variations of the rate ${T_{\rm d}}/{T_{\rm c}}$ as a function of
the frequency must be estimated to evaluate the second factor.

\begin{description}

\item[Dust thermal emission.]

For a wavelength of 240 $\mu $m the rate can be directly measured 
from the {\em COBE}-DIBE maps at $|b|>20^\circ $, taking 
$T$ as the total radiation, which includes diffuse and clouds emission; 
and $T^*$ as the radiation once the background is subtracted (as explained 
above), which is $T^*=T_{\rm c}-\langle T_{\rm c} \rangle $.
If we assume $\langle |\Delta T_{\rm c}|\rangle \sim \langle T_{\rm c} \rangle $, i.e.
the typical fluctuations in the cloud flux is as large as the average
of the flux, which is justified in a distribution of few single clouds
over $|b|>20^\circ $ with many regions having a negligible contribution, then 

\begin{equation}
1+\frac{T_{\rm dust,d}(240 \mu m)}{T_{\rm dust,c}(240 \mu m)}\sim 
\frac{\int _{|b|>20^\circ }d\Omega T(\Omega )}
{\int _{|b|>20^\circ }d\Omega |T^*(\Omega )|} = 3.3
\label{ratio240}
.\end{equation}

The outcome implies that 30\% of the radiation at 240 $\mu $m at
$|b|>20^\circ $ comes from clouds and 70\% from the diffuse 
interstellar medium.  This is just an estimation, but it is not a very 
bad approach. A comparison could be made with the rate from
the model by Cox, Kr\"ugel \& Mezger (1986) which gives a value
even greater: $\sim $ 4.5 in the Galactic disc.

Clouds are colder than the diffuse 
medium: between 6 and 15 K for molecular clouds and around 
16 K for diffuse medium according to Greenberg \& Li (1996).
Not all the infrared excess can be due to molecular clouds, but all
of these are generally colder than the diffuse medium of dust associated
with gas. Indeed,
infrared-excess clouds are peaks of column density of dust probably associated
with molecular gas of colder temperatures than the rest of the dust
(Reach, Wall \& Odegard 1998, Lagache et al. 1998).
Hence, the increase in emission from clouds is 10 or 15 times
greater than from the interstellar medium when we compare
240$\mu $m and the microwave region, $\nu <100$ GHz, and
it is dominated by clouds emission in the microwave region
(see, for example, Fig. 14 of Beichman 1987; or Fig. 1
of Cox, Kr\"ugel \& Mezger 1986) and

\begin{equation}
1+\frac{T_{\rm dust,d}(\nu <100\ {\rm GHz})}{T_{\rm dust,c}(\nu <100\ {\rm GHz})}
\approx 1
.\end{equation}

Thus, the second factor for thermal radiation for these 
microwave frequencies is

\begin{equation}
\left(\frac{\left(\frac{\Delta T_{\rm dust}(\nu <100\ {\rm GHz})}{T_{\rm dust}
(\nu <100\ {\rm GHz})}\right)}
{\left(\frac{\Delta T_{\rm dust}(240\ \mu{\rm m})}
{T_{\rm dust}(240\ \mu{\rm m})}\right)}\right)\sim 3.3
\label{vib2}
.\end{equation}

\item[Dust rotational emission.]
A factor must also be taken into account for the rotational
emission of the dust. First, an error in the  
function given by Draine \& Lazarian (1998a)
may be included in this factor since the paper uses different
 parameters that are poorly known.
Secondly, the rate between small particles and large particles is 
lower for clouds than for the diffuse interstellar medium
(Greenberg \& Li 1996) because there are processes that destroy small
grains (Puget, L\'eger \& Boulanger 1985; Puget \& L\'eger 1989), and
this would decrease the emission. Thirdly,
the density in the clouds is also higher, thereby providing higher emission.
There is not enough data in
the literature to carry out accurate calculations. In any case, my intention
is to show the order of magnitude of the Galactic contribution rather than
make accurate predictions, so I think it is enough to take the dust rotational
emission shape of Fig. \ref{Fig:factor1}, divided by $T_{dust}(240{\rm \mu m})$,
and multiply this by some 
free parameter $f$. The range of this parameter will be between 0 and 1,
so the intensity of the rotational emission is not overestimated.
It will be conservatively estimated in the range of possible values compatible with our
knowledge of clouds and rotational emission.
The factor has to be less than unity because the values
of the anisotropies predicted with a factor 1 exceed the total observational
 values for anisotropies measured by the {\em COBE}-DMR and other surveys.

\end{description}

Summing up, the second factor in the  microwave region
is taken to be 1 for synchrotron emission,
3.3 for thermal dust emission
and an unknown factor $0<f<1$ for rotational emission.
The calculation of the second factor corresponding to the total
dust emission is calculated for each frequency as a weighted average
of the thermal and rotational emissions whose weights are
the respective intensities of the first factors. 

\end{description}

\begin{figure}
\begin{center}
\mbox{\epsfig{file=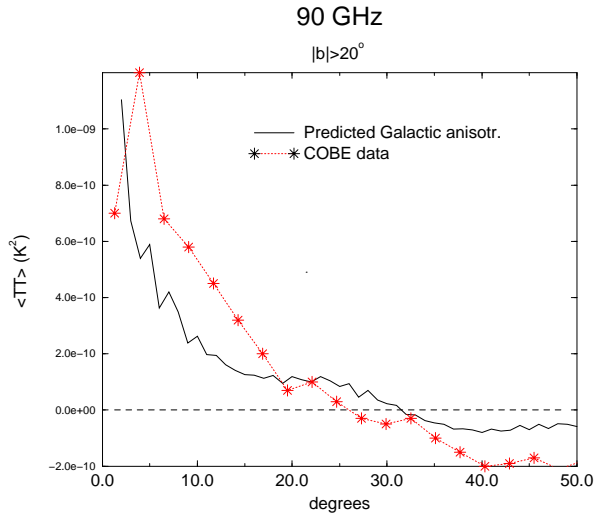,height=7cm}}
\end{center}
\caption{Predicted Galactic two-point correlations at 90 GHz, $|b|>20^\circ $,
according to eq. (\protect{\ref{corrtotal}}) compared with the {\em COBE}-DMR data
(Hinshaw et al. 1996).}
\label{Fig:corrtotal}
\end{figure}

So, with all this information the question to solve is how large are 
anisotropies due to the Galaxy at microwave frequencies. The answer is in
the eq. (\ref{corrtotal}). For example, at 90 GHz, the results are 
those shown in Fig. \ref{Fig:corrtotal}. Thermal emission of dust 
is the predominant one in this range. For this frequency, rotational emission with $f<1$ 
does not contribute significantly, so the exact value of the factor $f$ may 
be ignored. Synchrotron effects are also negligible since the relative
fluctuations in Fig. \ref{Fig:corr}a) are quite low.

As can be observed in Fig. \ref{Fig:corrtotal}, the amplitude of the
anisotropy is of the same order as that observed by {\em COBE}-DMR.
This casts doubts on the origin of MBRAs as, as will be discussed
in section \ref{.totGal}, they may be totally of Galactic origin.

The results from ULISSE experiment (Bernardis et al. 1992) or 
other previous experiments (Melchiorri et al. 1981) at 6$^\circ $ scale
show that CMBR anisotropies at 90-870 GHz or 410 GHz respectively
are less than 35 or 40 $\mu K$ at some regions at high 
Galactic latitudes, after Galactic emission subtraction. 
However, these authors use a Galactic emission model 
which does not take into account the cold molecular clouds discussed in this paper.
From eq. (\ref{corrtotal}), averaged over $|b|>20^\circ $,
$\sqrt{\langle TT\rangle (6^\circ ) (870 \ {\rm GHz})} \sim 50 \mu K$ (taking
the ``second factor'' as $f \sim 1$ since cold/diffuse clouds flux ratio varies
negligibily at these frequencies) and
$\sqrt{\langle TT\rangle (6^\circ ) (90 \ {\rm GHz})} \sim 19 \mu K$ ($f \sim 3.3$).
Fluctuations for the intermediate frequencies cannot be calculated
since the value of $f$ (second factor) as a
function of the frequency is unknown. Nevertheless, it can be 
seen that the order of magnitude of expected CMBR anisotropies and
those from the Galaxy as calculated in this paper are nearly the same.

\section{Is it possible that MBRAs are purely Galactic in origin rather than 
cosmological?}
\label{.totGal}

Maybe under some particular, but not impossible, conditions all the microwave background
radiation anisotropies be due to 
Galactic foregrounds rather than cosmological in origin.
Some arguments could be given against this and I will discuss them
in next subsections.

\subsection{Frequency dependence of the MBRAs amplitude}

The main argument in favour of a cosmological origin of the MBRAs is that 
these do not depend on frequency. Strictly speaking,
observations point out that the anisotropies at 53 and 90 GHz
are nearly the same, but that they are higher at 31 GHz (by a factor 2 or so
in difference, Hinshaw et al. 1996). This favours 
a cosmological origin of the MBRAs but not to the total exclusion of 
a Galactic origin.

The main argument against Galactic anisotropies with no
frequency dependence in the range between 50 and 90 GHz has been
that any emission---dust, free-free or synchrotron----gives
a dependence proportional to $\nu ^\alpha $ with $\alpha \ne 0$.
However, we should consider that there may be some range of
frequencies of transition between $\alpha > 0$---thermal dust
emission---and $\alpha <0 $---other kinds of emission.
The introduction of a quite potent rotational emission 
(see Fig. \ref{Fig:factor1}) is not normally taken into account 
and it is another important element, which has a negative
$\alpha $ for $\nu >20$ GHz. 
In the intermediate range between both regimes, the amplitude of
the anisotropies follows approximately a constant 
dependence with frequency, i.e. $\alpha \approx 0$.

\begin{figure}
\begin{center}
\mbox{\epsfig{file=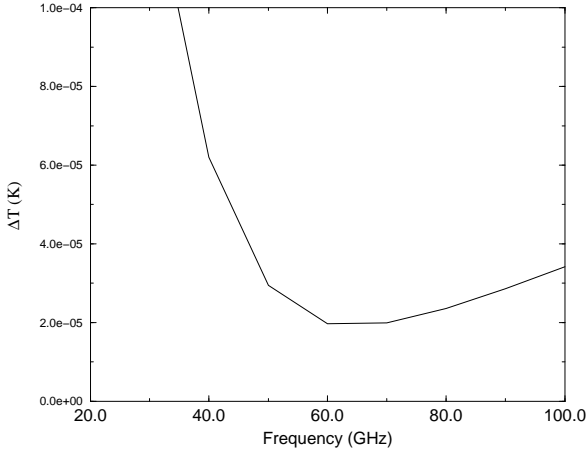,height=7cm}}
\end{center}
\caption{Multiplication of first and second factors of
eq. (\protect{\ref{corrtotal}}) for dust anisotropies
when the parameter $f$ is equal to 0.8. Normalized
such that $\Delta T(240 \mu m)=T(240 \mu m)$.}
\label{Fig:1_2_fact}
\end{figure}

Is it possible to get from Galactic sources of anisotropies the
same two-point correlation function for 53 and 90 GHz as observed
by {\em COBE}? The answer is yes: when $f\approx 0.8$
(see Fig. \ref{Fig:1_2_fact}). As observed in the figure, there is
a coincidence in the multiplication of the first and second factors
contained in  eq. (\ref{corrtotal}) for 53 and 90 GHz, so the amplitude of the correlation
will be the same. Synchrotron contribution is low.
There is also a dip around 60 GHz which separates slightly
from a flat spectrum but this range has not yet been observed accurately.
With the present model of rotational emission (model A of
Draine \& Lazarian 1998a), there is an excess of anisotropies
for 31.5 GHz with respect the {\em COBE}-DMR observations. This, however,
might be reduced when we choose other parameters for the
dust rotational emission predictions by Draine \& Lazarian (1998a).
The aim here has been to show that it is perfectly possible to explain
the observational anisotropies at microwave frequencies in terms
of Galactic clouds rather than constructing an accurate model.
The rotational emission is not well known and there is a lot
of work that remains to be done to get exact 
results. Up to now, only rough estimates can be carried out and
these indicate that, contrary to what was believed in the past,
there exists the possibility that the totality of the MBRAs are Galactic.

The main objection to this argument might be that it would
be an enormous coincidence that $f$ should be 0.8 or, equivalently, in
any rotational-emission prediction, that the range of nearly constant
$\alpha $ should be between 50 and 90 GHz. 

The predicted existence of rotational
emission much more intense than free-free emission (Draine \& Lazarian 1998a)
is controversial and not yet proven. Its existence is used in this paper as
a plausible candidate for a non-thermal contribution of emission correlated
with dust. However, it is not essential to justify anisotropies at 53 GHz.
Other mechanisms could be used instead. Free--free is an option, though 
there are problems in justifying the necessary flux for the total anisotropies. 
Another mechanism is the thermal
fluctuations of the magnetization within individual interstellar grains when 
most interstellar Fe exists in a moderately ferromagnetic material
(Draine \& Lazarian 1998b). This last mechanism might be nearly independent
of frequency in the 20-100 GHz range (see Fig. 7 of Draine \& Lazarian (1998b)
for Fe$_3$O$_4$ grains) avoiding the excess at 31.5 GHz described above.
At least, one form of contribution---free-free or rotational emission
or any other---is necessary to explain
the correlations between the anisotropies and far-infrared maps
(Kogut et al. 1996a; de Oliveira-Costa et al. 1997; de Oliveira-Costa et al. 1998). 
The role played by rotational emission with a spectral index $\alpha $
less than zero can be substituted by the free emission or others. 
It was shown that Galactic cirrus emission is high enough to explain observed anisotropies
at 90 GHz (Fig. \ref{Fig:corrtotal}). Whether this level of anisotropy
may be maintained for lower frequencies down to 50 GHz would be merely a question
of fitting some parameter to other emissions.
We can even fit the three or more emission types (free-free, dust rotational,
dust thermal emission, magnetic dipole emission from dust grains,...) 
at the same time in order to get a nearly flat 
continuum spectrum between 50 and 90 GHz, as well as to get a
conspicuous increase for lower frequencies.

\subsection{MBRA angular size}
\label{.angsize}

One remarkable feature of MBRAs that rouses suspicion about their
relationship to our Galaxy is the coincidence of the typical angular
size of their structures with the typical angular size of nearby clouds.
These structures have an appearance very similar
to the clouds observed in other frequencies.

\begin{figure}
\begin{center}
\mbox{\epsfig{file=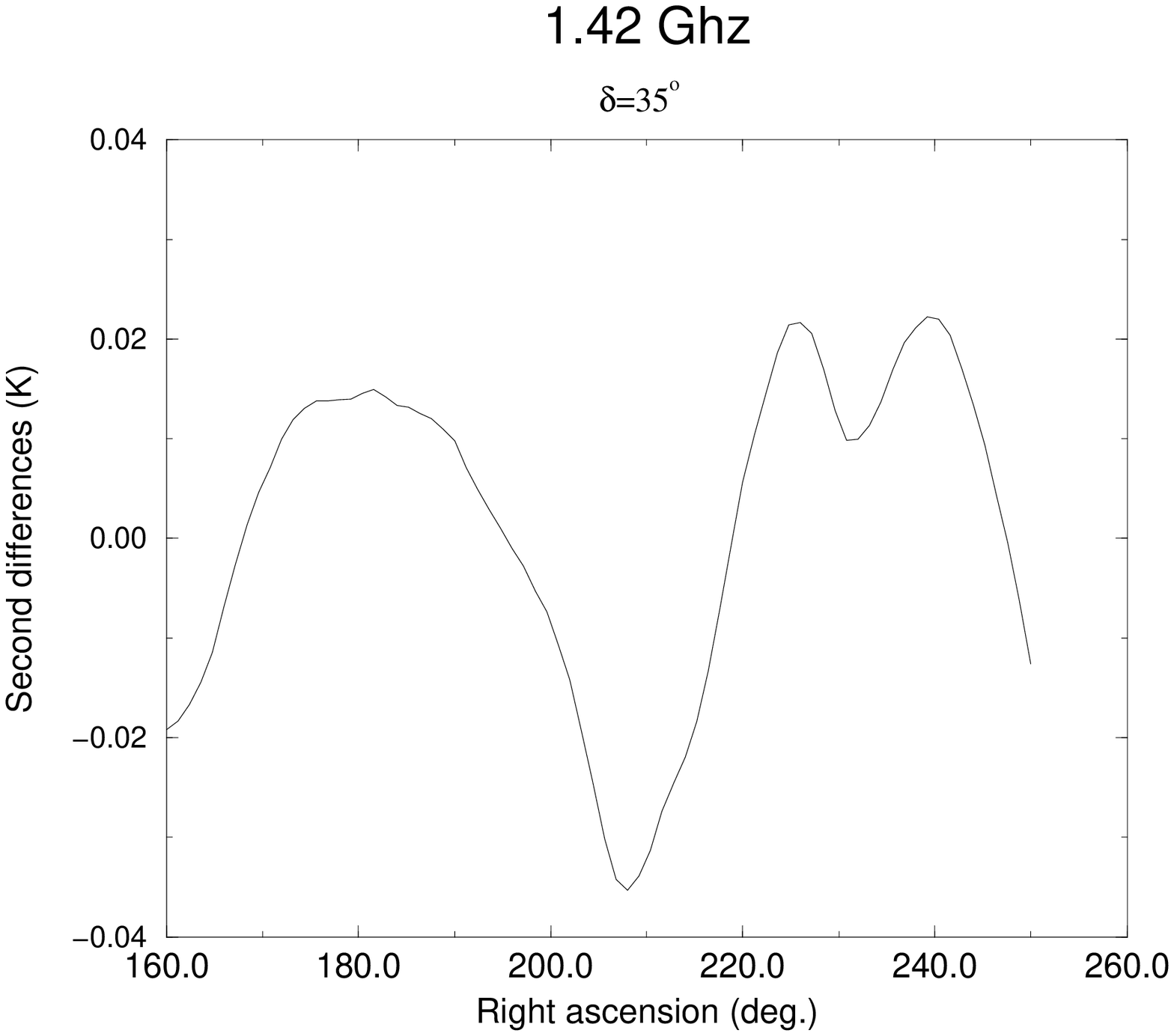,height=7cm}}

a)

\end{center}
\caption{Second differences of antenna temperature in different surveys
at $\delta =35^\circ $:
a) 1.42 GHz from the Max Planck Institute for Radio Astronomy (Reich 1982;
Reich \& Reich 1986); b) 4.85 GHz from NRAO (Condon, Broderick \& Seielstad 
1991; Condon, Griffith \& Wright 1993; Condon et al. 1994); 
c) 15 GHz from the Tenerife Experiment (Guti\'errez et al. 1997), solid line, 
and 53 and 90 GHz from {\em COBE}-DMR (Bunn, Hoffman \& Silk 1996), 
(Figure taken from Guti\'errez et al. 1997);
d) 240 $\mu $m (1250 GHz) from {\em COBE}-DIRBE 
(Boggess et al. 1992); e) 100 $\mu $m (3000 GHz) from {\em IRAS} (Wheelock 
et al. 1994).}
\label{Fig:secdiff}
\end{figure}

As an example, compare Fig. 4 of Guti\'errez et al. (1997,
reproduced in Fig. \ref{Fig:secdiff} c)), showing
structures observed by the  Tenerife Experiment, and Fig. \ref{Fig:secdiff}
a), b), d), e) for other frequencies. 
The second differences are evaluated according to 

\[
{\rm Second\ diff.}(\alpha, \delta)=T(\alpha, \delta ) -\frac{1}{2}
[T(\alpha+8.1^\circ /\cos(\delta ), \delta )
\]\begin{equation}+T(\alpha -8.1^\circ /\cos(\delta ), \delta )]
,\end{equation}
where $T$ is the antenna temperature of the radiation received in
a beam of FWHM=5 deg \ \footnote{To get the equivalent antenna 
temperature obtained from the Tenerife Experiment
with a FWHM=5$^\circ $ beam we do a convolution with a Gaussian
response with $\sigma =2.1^\circ $.}.

The aspect of the anisotropies is similar at all frequencies,
and the widths of the peaks are similar\footnote{
There is a blank strip in IRAS 100 $\mu $m map, which intercepts 
$\delta =35^\circ $ at $\alpha =173^\circ $. Thence, the downward peak 
of Figure \ref{Fig:secdiff} e)
near this right ascension should not be considered.}.
However, it is usually claimed that anisotropies between 
20 and 100 GHz are predominantly cosmological while the other
frequencies are dominated by the Galactic contribution.

\begin{figure}
\begin{center}
\mbox{\epsfig{file=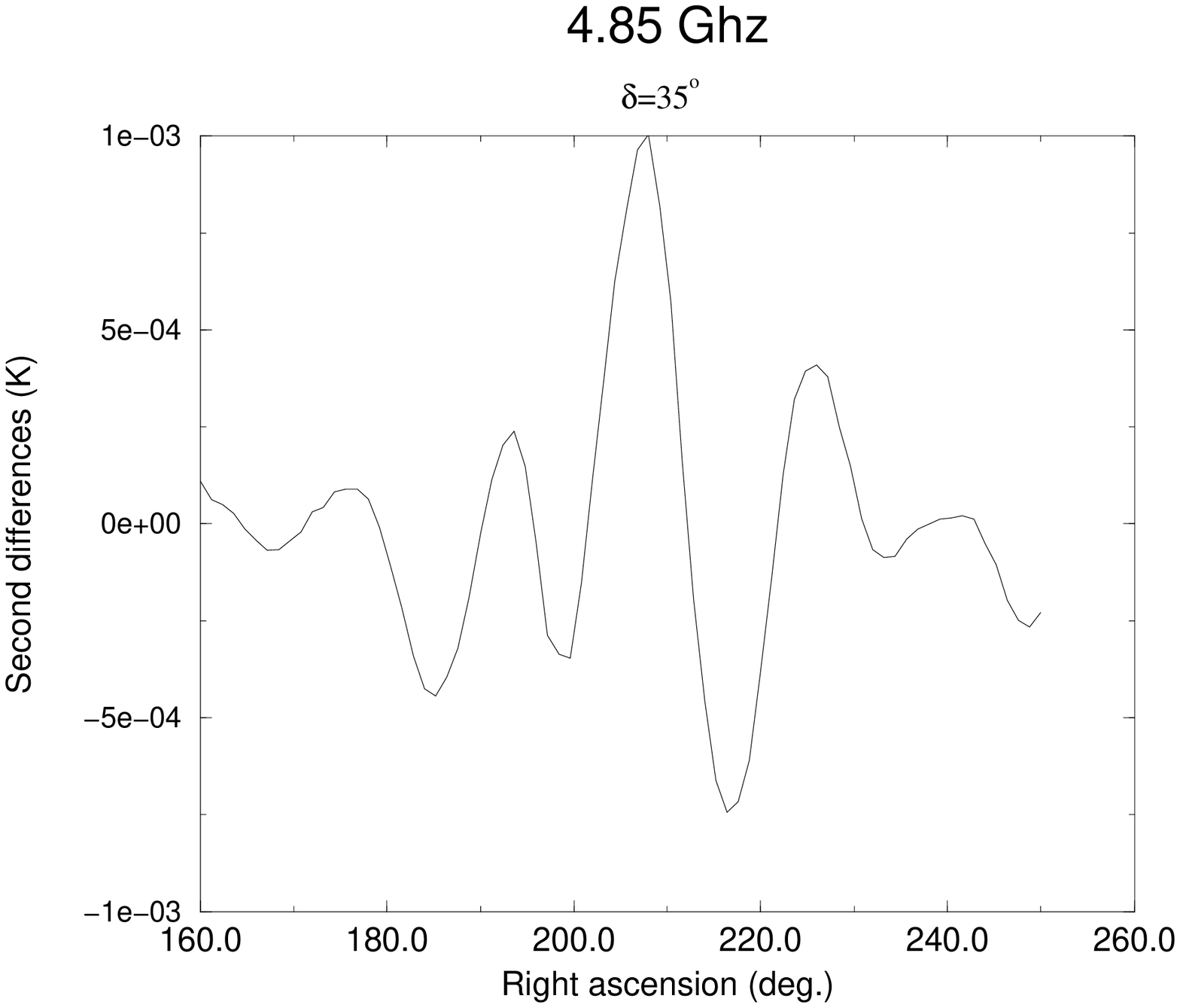,height=7cm}}

Fig. \ref{Fig:secdiff} b)
\end{center}
\end{figure}

\begin{figure}
\begin{center}
\mbox{\epsfig{file=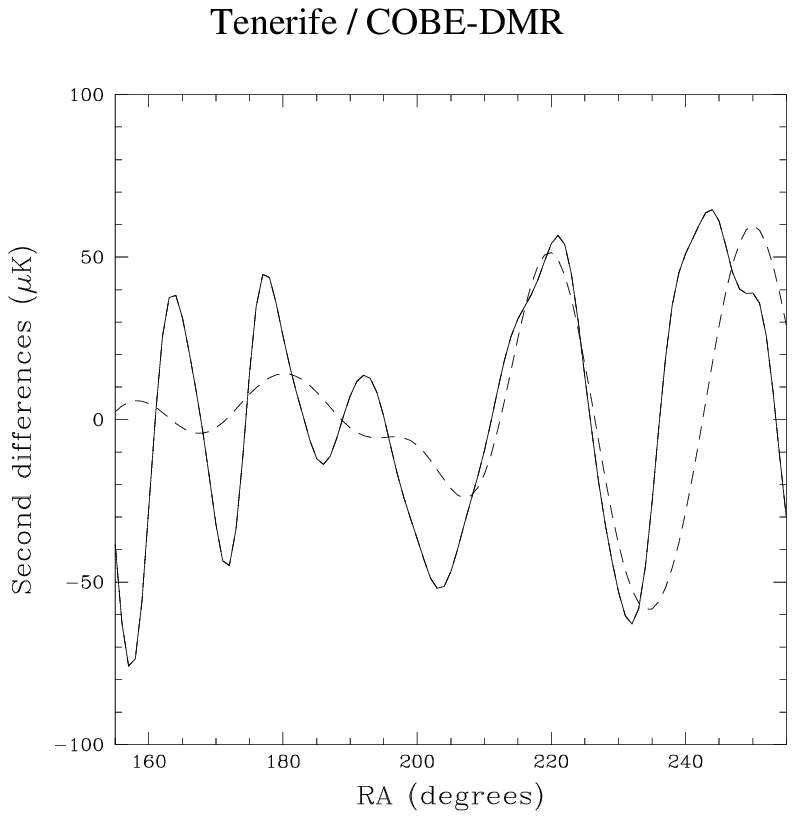,height=7cm}}

Fig. \ref{Fig:secdiff} c)
\end{center}
\end{figure}

\begin{figure}
\begin{center}
\mbox{\epsfig{file=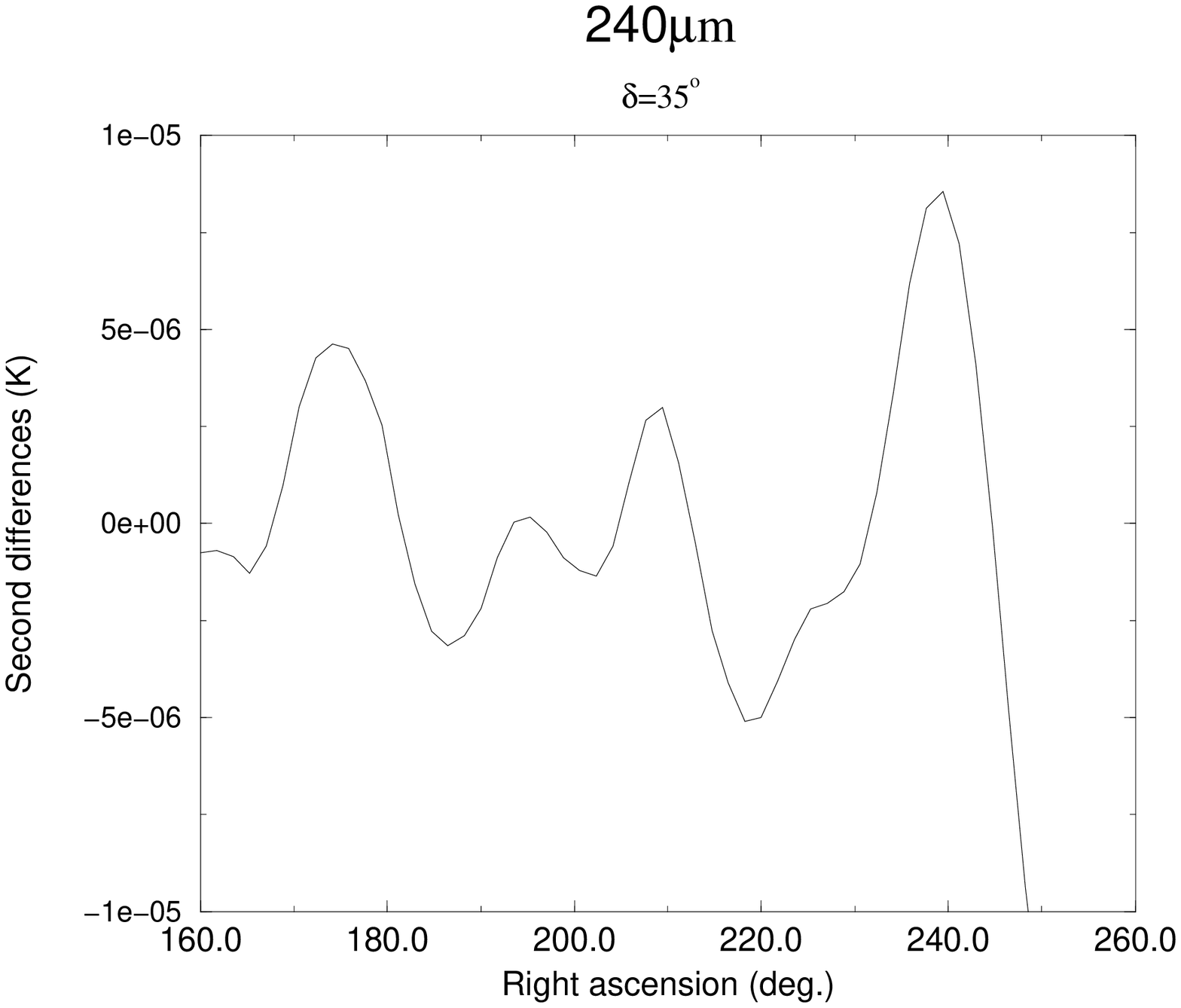,height=7cm}}

Fig. \ref{Fig:secdiff} d)
\end{center}
\end{figure}

\begin{figure}
\begin{center}
\mbox{\epsfig{file=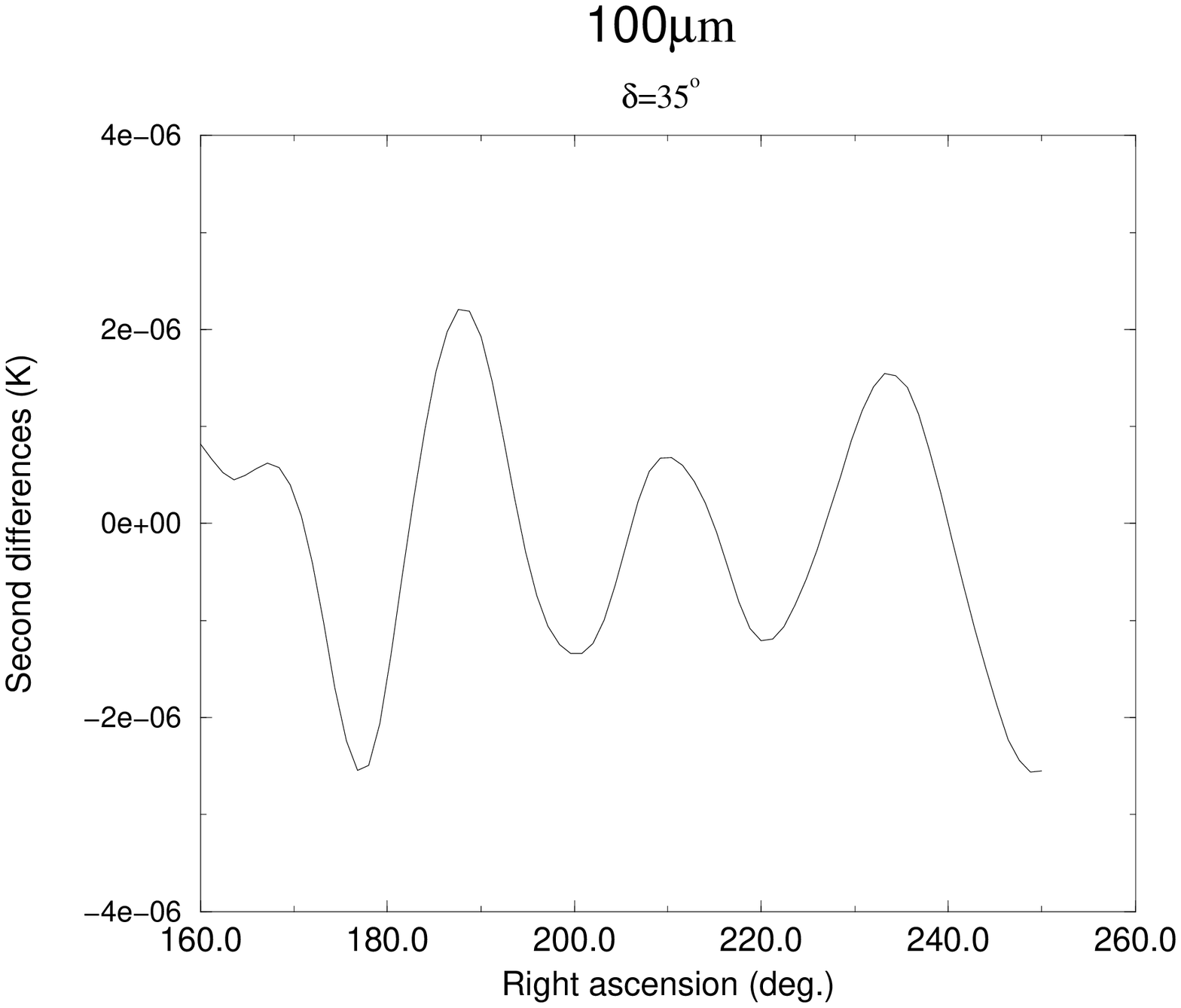,height=7cm}}

Fig. \ref{Fig:secdiff} e)
\end{center}
\end{figure}

We also see this likeness in Fig. \ref{Fig:corrtotal}. 
The correlation becomes null---first zero---around 30 deg at both 
Galactic and observed {\em COBE}-DMR. 
The first zero is related to the average angular
size of the clouds when they are responsible for the anisotropies
(L\'opez-Corredoira et al. 1998),
and this coincidence implies that the average angular size of the supposed
cosmological structures is approximately equal to that of the
nearby clouds giving rise to anisotropies.

The linear size of giant clouds is normally between 
20 and 100 pc (Scoville \& Sanders 1986) and can even 
reach sizes between 60 and 300 pc (Magnani, Blitz \& Mundi 1985; 
Heiles, Reach \& Boo 1988).
The projection of nearby structures such as these gives rise to
inhomogeneities and irregular arcs 
extending between 10$^\circ$ and 50$^\circ $ on the sky (Low \& Cutri 1994)
in the form of infrared cirrus.
In the range from $20$ and $100$ GHz, the Galactic contribution must
also provide anisotropies with a first zero around 20 or 30 deg
because this is the mean angular size of the clouds, independent
of the frequency (the first zero for synchrotron emission is somewhat less 
than 20 degrees as seen in Fig. \ref{Fig:corr}a).
This is precisely what is observed for the total anisotropies.
Most authors claim that these anisotropies are cosmological rather 
than Galactic but the coincidence referred to here might be due to more 
than just chance. Though this does not prove anything, it
is nevertheless a sign in favour of the Galactic predominance in MBRAs.

\subsection{Correlation among structures and the power spectrum
at different frequencies}
\label{.diffreq}

Although anisotropies over the whole range of the 
electromagnetic spectrum may be due to
inhomogeneities in the Galactic distribution of dust and gas, this
does not mean that the flux maxima and minima at different 
frequencies must occur at exactly the same coordinates.
Several effects---synchrotron, free-free, dust emission, etc.---are responsible 
for continuum emission but some effects  predominate over
others at different frequencies. Different effects
arise in different regions: synchrotron is higher
where magnetic fields are stronger, free-free radiation where a
warm ionized medium is present and dust where the coldest temperatures
are reached (Bennett et al. 1992). 
Peaks of dust emission are also to be observed at different
positions with different frequencies since cold or warm dust
are in different locations; small particles---which are dominant at microwave 
frequencies (Draine \& Lazarian 1998a)---and large 
particles---which are dominant in the far infrared 
(Greenberg \& Li 1996)---may be distributed differently in the clouds, etc. 
Thus, we cannot expect uniformity in Galactic structures
at different frequencies, i.e. an exact correlation for different
frequencies. Such a nonuniformity is observed, for instance, by 
Davies, Watson \& Guti\'errez (1996, their Fig. 10).

In any case, the correlation among different frequencies is not totally
null. Comparison of the different
plots in Fig. \ref{Fig:secdiff} shows a certain correlation. 
In Fig. \ref{Fig:corr_cross}, some cross-correlation 
is also observed at scales between 8 and 30
deg, while there is an anticorrelation for less than 8 deg.
The microwave continuum in the range between
14 and 90 GHz was also found to be correlated with 100$\mu $m thermal emission
from interstellar dust (Kogut et al. 1996a; de Oliveira-Costa et al. 1997;
de Oliveira-Costa et al. 1998). 
This was interpreted as a good correlation existing between dust 
and free-free emission (Kogut et al. 1996a, 1996b); however, the correlations
between H$\alpha $---normally a good tracer of free-free emission---with CMB
and DIRBE maps are weak (Leitch et al. 1997; Kogut 1997). Leitch et al. (1997)
alternatively proposed an anomalous bremsstrahlung emission from hot gas, but this
was again inconsistent with the observed power radiated (Draine \& Lazarian 1998a), 
so other kinds of emissions correlated with dust must be present.

Hence, it must be concluded that correlation among different structures
for different frequencies cannot be an
argument either for proving or disproving the Galactic origin of the anisotropies,
although some non-negligible Galactic contamination is necessary in order to
explain these last correlations.

The statistical distribution of the fluctuations given by the 
two-point correlation function, or its Fourier transform---the power
spectrum---is also expected to vary at different frequencies.
The coldest parts of the clouds produce dominant emission at lower
frequencies so the cloud shapes vary when the frequency varies.
Moreover, the  dust power spectrum
is very sensitive to the region of sky selected
(Guarini, Melchiorri \& Melchiorri 1995), the background subtraction
and other details.
As a matter of fact, different spectral indices have been obtained by different
authors for the dust emission: $-3.5 \le n \le -2.5$ (Gautier et al. 1992),
$-2.5 \le n \le -1.5$ (Melchiorri et al. 1996),
$n=-2.5$ for $|b|>45^\circ $ (Schlegel, Finkbeiner \& Davis 1998), etc.

At microwave frequencies, we cannot expect the same power spectra as
in the far infrared. Some unknown changes
are expected. According to this, power-spectrum
 or two-point correlation function shapes should not be used to
determine whether the anisotropies come from the Galaxy or are
cosmological. The fact that the angular power spectra observed at high 
galactic latitudes  by {\em COBE}-DIRBE are
steeper than the {\em COBE}-DMR power spectrum
should not be used, as in Wright (1998), to evaluate the
degree of Galactic contamination. 
The shape of the correlation
function is also different from the observational data in the microwave region
in Fig. \ref{Fig:corrtotal}:
there is a deficit of correlation at $\theta \sim 10$ deg and
an excess of correlation at $\theta \sim 25$ deg.
This may be, as has been said, because the shape of dust in 
far infrared was extrapolated in spite of the fact that variations were 
expected (\S \ref{.diffreq}), and that the cloud intensity fall-off
from its centre is different at 240 $\mu $m from that at microwave frequencies.

\subsubsection{The magnitude of the effects of the two-point correlation
function shape variation with frequency}
\label{.shapevar}

The extrapolation of dust emission as a combination of three factors
was carried out in (\ref{ext2}) under the assumption of a non-angular
dependence of the first and second factor. This dependence
would introduce  much more complex calculations for which we do not have 
 accurate enough data (it was not got sufficient information about the
correlations at wavelengths longer than 240 $\mu $m), although it can be
proven that the order of magnitude of the previous calculations would not
change. In any case, the approach taken here is better than any other prior
to this paper and the results are more trustworthy.

According to the hypothesis that
 clouds colder than the diffuse interstellar medium
produce the anisotropies, the variation of the temperature within these
clouds would produce variations in the $<TT>(\theta )$ shape.
In a simple model of emission, in which the cloud flux is proportional
to $\nu ^2 B(\nu, T_t)$, where $T_t$ is the effective grain temperature,
the antenna temperature $T$ is proportional to $B(\nu, T_t)$. $T_t$ is large
enough to take the Rayleigh--Jeans approximation 
in the microwave range, so $T\propto T_t$.
This implies that the variation of the antenna temperature in the microwave
region
is proportional to the variation of the effective grain temperatures.
For the calculation of $<TT>(\theta )$, our approach 
contains a relative error 

\begin{equation}
\frac{S(<TT>(\theta ))}{<TT>(\theta )}\approx 2\frac{\Delta T_t(\theta )}{<T_t>}
\label{errorTT}
,\end{equation}
where $\Delta T_t(\theta )$ is the mean variation of $T_t$ with respect its 
average value along two lines of sight separated by an angle $\theta $.
This is between $0.7$ and $1.3$ for a maximum relative variation of the
effective grain temperature depending of the angle of 15\%, which is quite a reasonable value
(for instance, clouds within a range of temperatures between 10 and 16 K
with a mean temperature of 13 K produce a $T_t(\theta )$ between 
13 K and 16 K, i.e. 14.5$\pm $ 1.5, a 10\%).
This may justify the difference of the shape in the plots of 
Figure \ref{Fig:corrtotal}. 
In any case, the mean amplitude over the whole range is the one calculated
above; there will be some angles in which the amplitude would be higher and
others in which it would be lower although the Galactic contamination 
is within the order of magnitude of the observed anisotropies.

\begin{figure} 
\begin{center}
\mbox{\epsfig{file=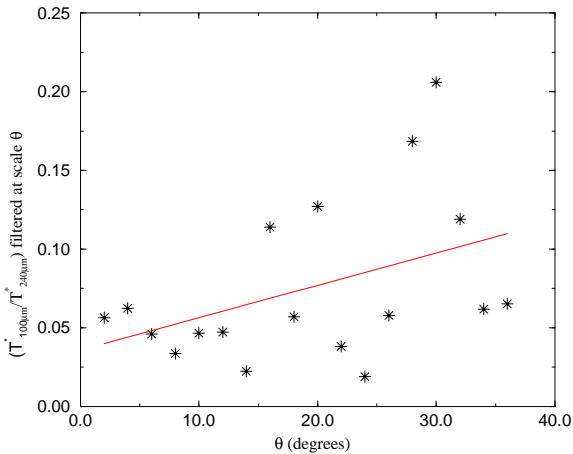,height=7cm}}
\end{center}
\caption{Ratio of flux excesses ($T^*$), i.e. after mean background subtraction,
between the maps of IRAS 100$\mu $m and DIRBE 240$\mu $ at $|b|>20^\circ $
with different filtering scale $\theta $. The solid line
is $0.036+0.0021 \ \theta $.
}
\label{Fig:c}
\end{figure}

From the IRAS 100$\mu $m and DIRBE 240$\mu $m maps, we derive roughly
(because the noise is quite high) a dependence between the ratio of flux
excesses and the filtering scale $\theta $ (Fig. \ref{Fig:c}):

\begin{equation}
Ratio =(0.036\pm 0.024)+(0.0021\pm 0.0011)\ \theta
\end{equation}

Since this ratio as a function of the filtering scale
is equal to $\frac{1}{A}\int f(\theta )\ dA$, where $A=\pi \theta ^2$
and $f(\theta )$ is the ratio as a function of the correlation angle,

\begin{equation}
f(\theta ) =(0.036\pm 0.024)+(0.0031\pm 0.0016)\ \theta
\label{f1}
.\end{equation}

On the other hand,

\[
f(\theta )=\frac{T_{100 \ {\rm \mu m}}^*}{T_{240 \ {\rm \mu m}}^*}=
\frac{\tau (\nu _1^2 B(\nu _1, T_t(\theta ))\frac{c^2}{2k\nu _1^2}}
{\tau (\nu _2^2 B(\nu _2, T_t(\theta ))\frac{c^2}{2k\nu _2^2}}
\]\begin{equation}=
\frac{\frac{2h\nu _2^3}{c^2}\frac{1}{e^{\frac{h\nu _1}{kT_t(\theta )}}-1}}
{\frac{2h\nu _1^3}{c^2}\frac{1}{e^{\frac{h\nu _2}{kT_t(\theta )}}-1}}
\label{f2}
,\end{equation}
where $\nu _1$ and $\nu _2$ are respectively the frequencies corresponding
to 100 $\mu $m and 240 $\mu $m. From (\ref{f1}) and (\ref{f2}), we
derive a linear dependence (correlation coefficient with a linear fit:
0.9958):

\begin{equation}
T_t (\theta )=14.5+0.129\ \theta
.\end{equation}

This means that the range of temperatures for angles $\theta $
less than 30 degrees
is within $T_t=16.4\pm 1.9$ K, i.e. a relative error of 11.5\% and
the mean temperature of the clouds would be $\langle T_t\rangle =14.5$ K
(more or less in agreement with the calculations by Lagache et al. 1998).
If we use equation (\ref{errorTT}), this leads to a maximum error of
$<TT>(\theta )$ of 23\%. This may justify the difference 
of the shape in the plots of 
Figure \ref{Fig:corrtotal}, and it does not change the order of magnitude of
our calculations. An exact calculation is not carried out since the
uncertanties derived from Fig. \ref{Fig:c} are too much and the use
of a simple model of dust ($\nu ^2B(\nu, T)$) may be not totally correct to 
extrapolate $T^*$ from the given frequencies to the microwaves
(for instance, very small particles may contribute less than a 10\% of
the total flux (Greenberg \& Li 1996) which is negligible for our required accuracy). 
In any case, this was just to estimate the relative variation with $\theta $ 
and not to calculate the exact value of $T_t$.

\subsection{MBRA galactic latitude dependence}
\label{.posdep}

\begin{figure}
\begin{center}
\mbox{\epsfig{file=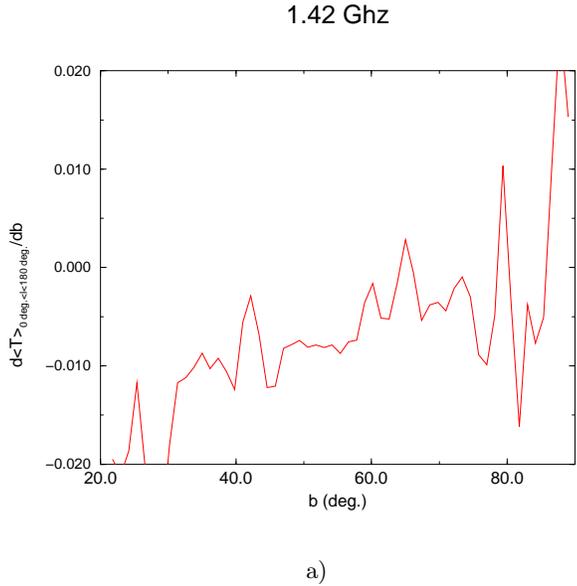,height=7cm}}

a)
\end{center}
\caption{Flux derivative versus Galactic latitude ($b$).
a) 1.42 GHz; b) 240 $\mu $m. }
\label{Fig:posdep}
\end{figure}

From Fig. \ref{Fig:posdep}, it is observed that there is some
positional dependence of the fluctuations. In this figure,
the flux derivative is shown instead of the flux because
it allows the fluctuations to be seen more clearly. The variation
of the fluctuations at the Galactic poles is something different
from that at intermediate latitudes.

\begin{figure}
\begin{center}
\mbox{\epsfig{file=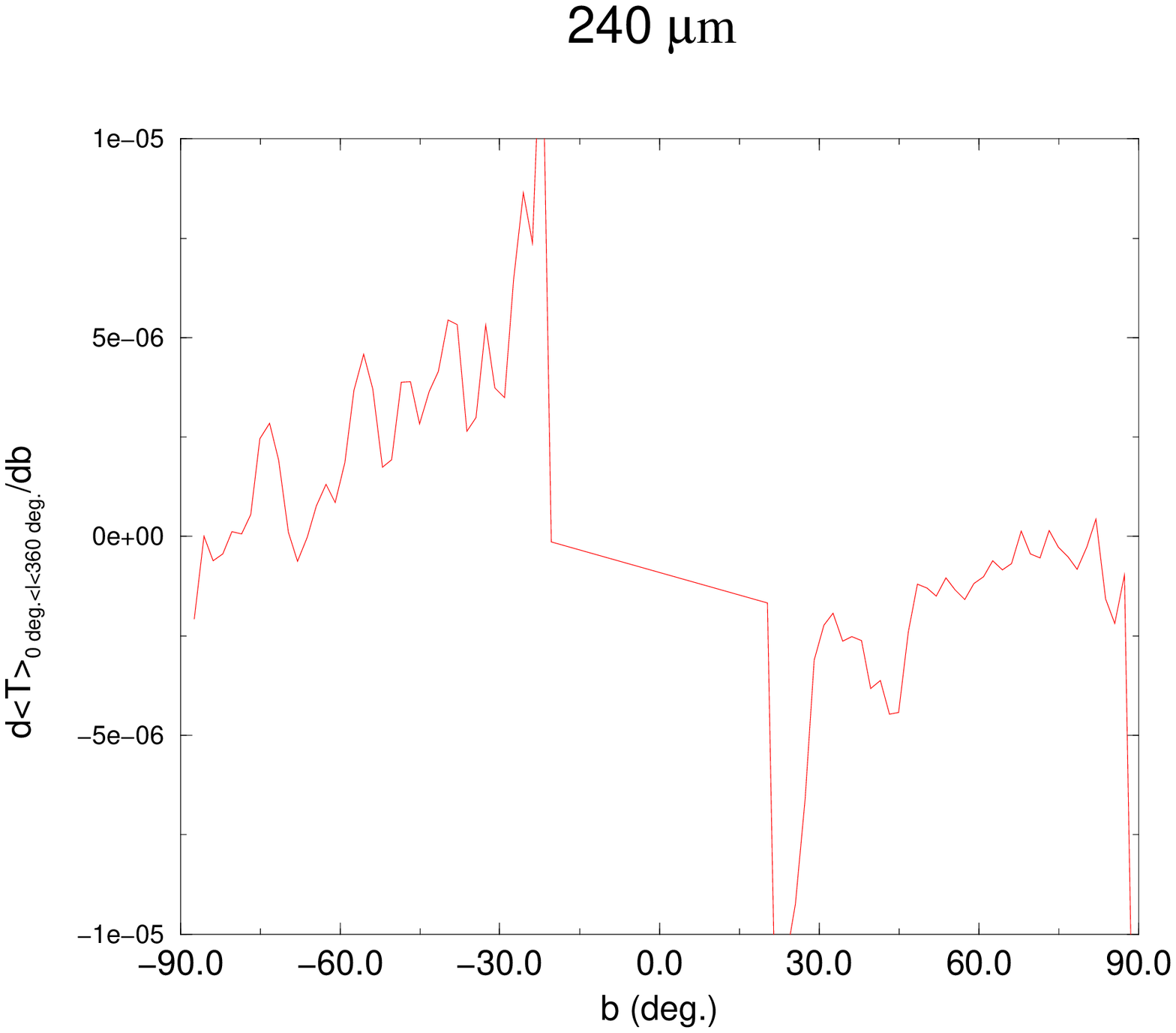,height=7cm}}

Fig. \ref{Fig:posdep} b)
\end{center}
\end{figure}

The density distribution of Galactic clouds
is irregular with a higher concentration in the plane
and a fall-off which can  be roughly  represented by an exponential
with a scale height around 75 pc (Scoville \& Sanders 1986). 
Clouds are distributed  very close  indeed to the plane. As a matter
of fact, nearly all clouds are less than 20 deg from
the plane. The remaining clouds that we observe are a few local
ones belonging to the plane and are very close to the Sun (Blitz 1991);
they are distributed randomly.
Any distribution in which the density depends only on $z$, the distance
to the plane, should give a column density proportional to $cosec(b)$,
but in this case the irregularity makes the density depend also on
the other two coordinates, and the column density follows another dependence
with respect to $b$. The number of clouds is  too small to
provide good statistics, so not a lot more can be said about this dependence.
Figure \ref{Fig:posdep} may show very slight trends but these are not too clear.

\begin{figure} 
\begin{center}
\mbox{\epsfig{file=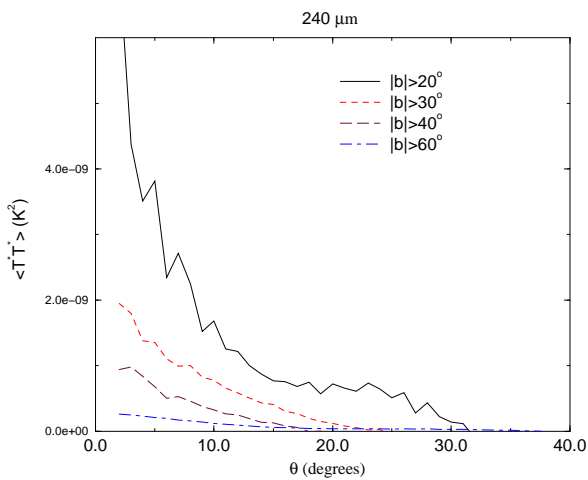,height=7cm}}
\end{center}
\caption{Angular correlation function at 240 $\mu $m.}
\label{Fig:corr240ang}
\end{figure}

If we do all the calculations of this paper in different regions,
for instance: $|b|>30^\circ $, $|b|>40^\circ $ and $|b|>60^\circ $, the
correlations expected would be somewhat lower. The anisotropies
from dust are shown in Fig. \ref{Fig:corr240ang}. Instead of (\ref{dust>20})
as first factor, the approximate fits derived from Reach et al. (1995) are
(values for $|b|>40^\circ $ were calculated by interpolation):

\[
T_{\rm dust}(\nu ,|b|>30^\circ )=\left(\frac{\nu }{900\ {\rm GHz}}\right)^2 
[2.11\times 10^{-5}
\]\[\times
B(\nu ,T=17.70\ {\rm K})+1.45\times 10^{-4}
B(\nu ,T=7.02\ {\rm K})]
\]\begin{equation}\times
\frac{c^2}{2k\nu ^2} \ \ {\rm K}
,\end{equation}

\[
T_{\rm dust}(\nu ,|b|>40^\circ )=\left(\frac{\nu }{900\ {\rm GHz}}\right)^2 
[1.68\times 10^{-5}
\]\[\times
B(\nu ,T=17.75\ {\rm K})+1.24\times 10^{-4}
B(\nu ,T=7.23\ {\rm K})]
\]\begin{equation}\times
\frac{c^2}{2k\nu ^2} \ \ {\rm K}
,\end{equation}

\[
T_{\rm dust}(\nu ,|b|>60^\circ )=\left(\frac{\nu }{900\ {\rm GHz}}\right)^2 
[1.26\times 10^{-5}
\]\[\times
B(\nu ,T=17.80\ {\rm K})+1.03\times 10^{-4}
B(\nu ,T=7.44\ {\rm K})]
\]\begin{equation}\times
\frac{c^2}{2k\nu ^2} \ \ {\rm K}
,\end{equation}

\begin{figure} 
\begin{center}
\mbox{\epsfig{file=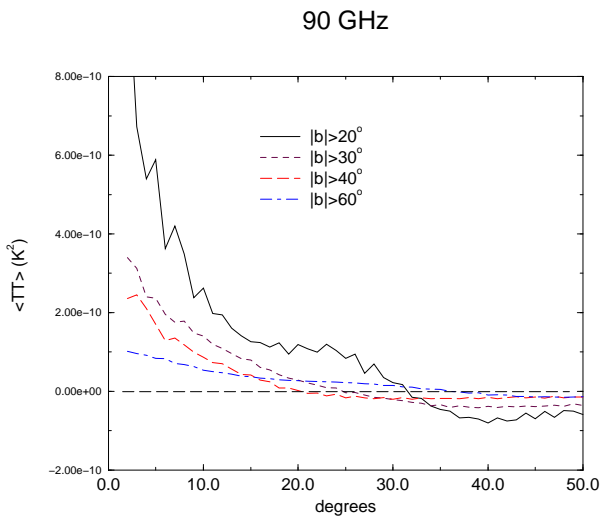,height=7cm}}
\end{center}
\caption{Predicted Galactic two-point correlations at 90 GHz at different zones.}
\label{Fig:corrtotalang}
\end{figure}

And the values for the second factors due to dust thermal emission, calculated
as above, are 3.5, 4.0 and 4.8, respectively, for 
$|b|>30^\circ $, $|b|>40^\circ $ and $|b|>60^\circ $.
At 90 GHz, rotational dust emission and 
synchrotron fluctuations are negligible, and those from the dust
thermal emission, derived according to (\ref{ext2}), predict a lower amplitude
of the fluctuations at higher latitudes, as is shown in Fig. 
\ref{Fig:corrtotalang}.

The fluctuations, proportional to the square root of $\langle TT\rangle $,
are about twice as high for $|b|>20^\circ $ than for $|b|>60^\circ $ at a
scale of around $\theta =5^\circ $.
Therefore, a decreasing fluctuation at higher latitudes is expected
if the Galactic emission was the sole or at least an important
contributor to the anisotropies. Perhaps the observed factor between
$|b|>60^\circ $ and $|b|>20^\circ $
is not exactly two, since this quantity
is subject to the errors of the calculations of the first and second factor,
but the order of magnitude is something of this order and there is likely to
exist a decreasing correlation with Galactic latitude.

In fact, MBRAs at microwave frequencies show a slightly decreasing
dependence on $|b|$ with the fluctuations (Smoot et al. 1992, Fig. 2).
They show that the correlations for $|b|>30^\circ $ and $|b|>40^\circ $
are nearly a half that for $|b|>20^\circ $ for angles $\theta $
less than 10$^\circ $, although their first-year
{\em COBE}-DMR data are quite noisy\footnote{I do not undesrtand
why these correlations were not calculated again with the four-year
{\em COBE}-DMR data but only for $|b|>20^\circ $.}. 
Data showing this dependence more accurately are still awaited, specially
when next experiments (PLANCK or MAP) be working.
Until now, it cannot be said what exactly the $b$ dependence 
in the anisotropies at 90 GHz is.

Another conspicuous dependence is that the anisotropies
in the southern Galactic hemisphere are higher than in the northern Galactic
hemisphere. This is still not a proof against the Galactic predominance
of anisotropies at microwave frequencies because the {\em COBE}-DMR survey also
observes higher anisotropies in the southern  Galactic hemisphere
(see Fig. 1 of Bennett et al. 1996)\footnote{I do not understand
why this fact is not commented on in that
paper, since the figure shows the difference of
both hemispheres quite clearly.
This difference is attributed to Galactic contamination, but 
may not  Galactic contamination be responsible 100\% of the anisotropies?}.

From this, the conclusion is that there is not a qualitative 
difference between the cosmological
anisotropies and the cloud anisotropies. The position dependence of
the MBRA is qualitatively similar in both microwaves and far infrared,
so this may not be a proof neither for
the cosmological origin nor for the Galactic origin. 

\subsection{Conclusions about possible Galactic predominance in
MBRAs anisotropies}

The conclusion is that under some particular, but not impossible, conditions,
all the microwave background
radiation anisotropies may be due to 
Galactic foregrounds rather than cosmological in origin.
There are no arguments yet to exclude this possibility although
this is not yet proved and the question remains open.
A testable prediction of such a case would be that the amplitude of the
fluctuations for $|b|>60^\circ $ would be about the half of that 
for $|b|>20^\circ $ for angles around 5 degrees.

The implications of such a question are extremely important, not merely for
refining some quantity or other or for making
certain corrections to get an accurate result for an individual parameter,
but because it would result in a different qualitative description of the 
Universe. The implications for inflation theories or the formation
of the large-scale structure of the Universe would be enormous,
and our ideas regarding such formations would change completely.
Hence, I think studying the influence of the Galaxy is a valuable exercise,
in order to avoid the hazarding of cosmological theories based
on cumulative errors, in which this paper claim to be still an open question.

\section{Conclusions}

The following main conclusions may be drawn from this paper:

\begin{itemize}

\item The extrapolation of anisotropies following the mean dust emission is 
a bad approximation since it does not take into account
the growing contrast of colder clouds in the background
of the diffuse interstellar medium. By considering
this effect, it is found that dust thermal emission anisotropies are 
higher than expected by other authors, and that their amplitude is comparable
to the observational data at 90 GHz.

\item Our ignorance of the different emission mechanisms 
around 50 GHz (free-free, dust rotational emission, magnetic dipole emission 
from dust grains) do not allow the firm
conclusion that anisotropies due to Galactic emission are not frequency
dependent but this possibility remains open.

\item If Galaxy-induced anisotropies are not 
responsible for the totality of the correlations
they would at least be a non-negligible part of them, so untrustworthy
cosmological conclusions could be reached from microwave background
radiation anisotropy analysis unless possible Galactic emission processes
are correctly subtracted.

\item If the Galaxy-induced anisotropies made up the total correlations
at microwave frequencies, then inflation, models of Galaxy formation and
many parts of the standard cosmology would be wrong. This is not 
impossible, though there is no firm evidence either for or against it as yet.

\end{itemize}

\begin{acknowledgements}

I acknowledge gratefully the comments and suggestions to improve the
content of the paper made by C. M. Guti\'errez, J. Lequeux,
F. Melchiorri and A. Abergel. 

\end{acknowledgements}


\begin{thebibliography}{99}

\bibitem{} Banday A. J., Wolfendale A. W., 1991, MNRAS 252, 462

\bibitem{} Beichman C. A., 1987, ARA\&A 25, 521

\bibitem{} Bennett C. L., Banday A. J., G\'orski K. M., et al.,
1996, ApJ 464, L1

\bibitem{} Bennett C. L., Smoot G. F., Hinshaw G., et al., 1992,
ApJ 396, L7

\bibitem{} Bernardis P. de, Masi S., Melchiorri F., Melchiorri B.,
Vittorio N., 1992, ApJ 396, L57

\bibitem{} Blitz L., 1991, in: Molecular clouds, R. A. James,
T. J. Milla (eds.), Cambridge University Press, Cambridge, p.\ 49 

\bibitem{} Boggess N. W., Mather J. C., Weiss R., et al., 1992, ApJ 397, 420

\bibitem{} Boulanger F., Abergel A., Bernard J. P., et al., 1996, A\&A 312, 256

\bibitem{} Boulanger F., Baud B., van Albada T., 1985, A\&A 144, 9

\bibitem{} Boulanger F., P\'erault M., 1988, ApJ 330, 964

\bibitem{} Burton W. B., Deul E. R., 1987, in: The Galaxy,
G. Gilmore, B. Carswell, eds., Reidel, Dordrecht, p.\ 141

\bibitem{} Bunn E. F., Hoffman Y. \& Silk J., 1996, ApJ 464, 1

\bibitem{} Celebonovic V., Samurovic S., Cirkovic M. M., 1997,
Publ. Astron. Obs. Belgrade 57, 105

\bibitem{} Combes F., Pfenninger D., 1997, A\&A 327, 453

\bibitem{} Condon J. J., Broderick J. J., Seielstad G. A., 1991, AJ 102, 2041

\bibitem{} Condon J. J., Giffith M. R., Wright A. E., 1993, AJ 106, 1095

\bibitem{} Condon J. J., Broderick J. J., Seielstad G. A., Douglas K.,
Gregory P. C., 1994, AJ 107, 1829

\bibitem{} Cox P., Kr\"ugel E., Mezger P. G., 1986, A\&A 155, 380

\bibitem{} Davies R. D., Watson R. A., Guti\'errez C. M., 1996, MNRAS 278,
925

\bibitem{} de Oliveira-Costa A., Kogut A., Devlin M. J., et al., 1997, ApJ
482, L17

\bibitem{} de Oliveira-Costa A., Tegmark M., Page L. A., Boughn S. P., 1998,
ApJ 509, L9

\bibitem{} Draine B. T., 1994, in: The Infrared Cirrus and Diffuse
Interstellar Clouds, ASP Conference Series 58,
R. Cutri, W. B. Latters, eds., San Francisco, p.\ 227

\bibitem{} Draine B. T., Lazarian A., 1998a,
ApJ 494, L19

\bibitem{} Draine B. T., Lazarian A., 1998b, preprint astro-ph/9807009

\bibitem{} Draine B. T., Lee H. M., 1984, ApJ 285, 89

\bibitem{} Femen\'\i a B., Rebolo R., Guti\'errez C. M., Limon M.,
Piccirillo L., 1998, ApJ 498, 117

\bibitem{} Frisch P. C., York D. G., 1986, in: The Galaxy and
the Solar System, R. Smoluchowski, J. N. Bahcall, M. S. Matthews, eds.,
University of Arizona Press, Tucson, p.\ 83

\bibitem{} Fukushige T., Makino J., Ebisuzaki T., 1994,
ApJ 436, L107

\bibitem{} Gautier T. N. III, Boulanger F., Perault M., Puget J. L., 1992,
AJ 103, 1313

\bibitem{} Greenberg J. M., Li A., 1996, in: New Extragalactic Perspectives
in the New South Africa, D. L. Block, J. M. Greensberg, eds.,
Kluwer, Dordrecht, p.\ 118

\bibitem{} Guarini G., Melchiorri B., Melchiorri F., 1995,
ApJ 442, 23

\bibitem{} Guti\'errez C. M., Hancock S., Davies R. D., et al., 1997,
ApJ 480, L83

\bibitem{} Heiles C., 1976, ApJ 204, 379

\bibitem{} Heiles C., Reach W. T., Koo B. C. 1988, ApJ 332, 313

\bibitem{} Hinshaw G., Banday A. J., Bennett C. L., et al., 1996,
ApJ 464, L25

\bibitem{} Knox L., Scoccimarro R., Dodelson S., 1998, 
preprint astro-ph/9805012

\bibitem{} Kogut A., 1997, AJ 114, 1127

\bibitem{} Kogut A., Banday A. J., Bennett C. L., et al., 1996a, ApJ 460, 1

\bibitem{} Kogut A., Banday A. J., Bennett C. L., et al., 1996b, ApJ 464, L5

\bibitem{} Lagache G., Abergel A., Boulanger F., Puget J.-L., 1998,
A\&A 333, 709

\bibitem{} Leitch E. M., Readhead A. C. S., Pearson T. J., Myers S. T., 
1997, ApJ 486, L23

\bibitem{} L\'opez-Corredoira M., Garz\'on F.,
Hammersley P. L., Mahoney T. J., 1998, MNRAS 301, 289

\bibitem{} Low F. J., Cutri R. M., 1994, Infrared Phys. Technol. 35, 291

\bibitem{} Magnani L., Blitz L., Mundi L., 1985, ApJ 295, 402

\bibitem{} Masi S., Calisse P., de Bernardis P., et al., 1990, in:
The Galactic and Extragalactic Background Radiation, IAU Symp. 139,
S. Bowyer, C. Leinert, eds., Kluwer, Dordrecht

\bibitem{} Mathis J. S., 1990, ARA\&A 28, 37

\bibitem{} Matsumoto T., Hayakawa S., Matsuo H., et al., 1988, ApJ 329, 567

\bibitem{} Meinhold P., Clapp a., Devlin M., et al., 1993, ApJ 409, L1

\bibitem{} Melchiorri F., Guarini G., Melchiorri B., Signore M., 1996,
ApJ 464, 18

\bibitem{} Melchiorri F., Melchiorri B. O., Ceccarelli C., Pietranera L.,
1981, ApJ 250, L1

\bibitem{} Pando J., Valls-Gabaud D. \& Fang L.-Z., 1998, Phys. Rev.
Lett. 81(21), 4568

\bibitem{} Pfenninger D., Combes F., 1994,
A\&A 285, 94

\bibitem{} Pfenninger D., Combes F., Martinet L., 1994,
A\&A 285, 79

\bibitem{} Puget J. L., Abergel A., Bernard J. P., et al., 1996, A\&A 308, L5

\bibitem{} Puget J. L., L\'eger A., 1989, ARA\&A 27, 161

\bibitem{} Puget J. L., L\'eger A., Boulanger F., 1985, A\&A 142, L19

\bibitem{} Reach W. T., Dwek E., Fixsen D. J., et al., 1995, ApJ
451, 188

\bibitem{} Reach W. T., Wall W. F., Odegard N., 1998, ApJ 507, 507

\bibitem{} Readhead A. C. S., Lawrence C. R., 1992, ARA\&A 30, 653

\bibitem{} Reich W., 1982, A\&AS 48, 219

\bibitem{} Reich P., Reich W., 1986, A\&AS 63, 205

\bibitem{} Reynolds R. J., 1991, ApJ 372, L17

\bibitem{} Schaefer J., 1994, A\&A 284, 1015

\bibitem{} Schaefer J., 1996, Europhys. Lett. 34(1), 69

\bibitem{} Schlegel D. J., Finkbeiner D. P., Davis M., 1998, ApJ 500, 525

\bibitem{} Schloerb F. P., Snell R. L., Schwartz P. R., 1987,
ApJ 319, 426

\bibitem{} Scoville N. Z., Sanders D. B., 1986, in: The Galaxy and
the Solar System, R. Smoluchowski, J. N. Bahcall, M. S. Matthews, eds.,
University of Arizona Press, Tucson, p.\ 69

\bibitem{} Smoot G. F., 1998, preprint astro-ph/9801121

\bibitem{} Smoot G. F., Bennett C. L., Kogut A., et al., 1992, ApJ 396, L1

\bibitem{} Sodroski T. J., Odegard N., Arendt R. G., et al., 1997,
ApJ 480, 173

\bibitem{} Suginohara M., Suginohara T., Spergel D. N., 1998,
ApJ 495, 511

\bibitem{} Tegmark M., 1998, ApJ 502, 1

\bibitem{} Wheelock S. L., Gautier T. N., Chillemi J., et al., 
1994, IRAS Sky Survey Atlas: Explanatory
Supplement, Jet Propulsion Laboratory, Pasadena

\bibitem{} White M., Scott D., Silk J., 1994, ARA\&A 32, 319

\bibitem{} Wright E. L., 1993, in: Back to the Galaxy, AIP Conf.
Proc. 278, S. S. Holt, F. Verter, eds., AIP, New York, p.\ 193

\bibitem{} Wright E. L., 1998, ApJ 496, 1

\bibitem{} Wright E. L., Mather J. C., Bennett C. L., et al., 1991,
ApJ 381, 200 


\end{thebibliography}
\end{document}